\documentclass{article}

\usepackage{moreverb,url}
\usepackage[colorlinks,bookmarksopen,bookmarksnumbered,citecolor=red,urlcolor=red]{hyperref}

\usepackage{amssymb}
\usepackage{xcolor}
\usepackage{soul}
\usepackage{MnSymbol}
\usepackage{xfrac}
\usepackage{siunitx}
\usepackage[utf8]{inputenc}
\usepackage{setspace}
\usepackage{amsmath}
\usepackage{amsfonts}
\usepackage{mathtools}
\usepackage{tikz}
\usetikzlibrary{positioning,calc,shapes,arrows,shadows,fit,scopes}
\usetikzlibrary{backgrounds}

\usepackage{natbib}
\usepackage{authblk}

\usepackage{tabularx}
\usepackage{makecell}

\usepackage{pgfplots}
\pgfdeclarelayer{background}
\pgfsetlayers{background,main}

\usepackage{xifthen}

\usepackage{esdiff}
\usepackage[noabbrev]{cleveref}
\usepackage{ifthen}

\usepackage{esdiff}

\usepackage{standalone}

\usepackage{MnSymbol} 
\usepackage{booktabs}
\usepackage{nicefrac}

\usepackage{graphicx}
\usepackage{subcaption}

\usepackage{listings}
\usepackage[ruled,lined,algo2e]{algo2e}

\usepackage[normalem]{ulem}
\usepackage{booktabs}
\usepackage{multirow}
\usepackage{relsize}

\usepackage{algorithm}

\SetCommentSty{mycommfont}

\usepackage{array}
\newcolumntype{L}[1]{>{\raggedright\let\newline\\\arraybackslash\hspace{0pt}}m{#1}}
\newcolumntype{C}[1]{>{\centering\let\newline\\\arraybackslash\hspace{0pt}}m{#1}}
\newcolumntype{R}[1]{>{\raggedleft\let\newline\\\arraybackslash\hspace{0pt}}m{#1}}

\usepackage[acronym,nogroupskip,nomain,indexonlyfirst,toc]{glossaries}
\glsdisablehyper
\setacronymstyle{long-short}
\newacronym{hpc}{HPC}{high performance computing}
\newacronym{lbm}{LBM}{lattice Boltzmann method}
\newacronym{cfd}{CFD}{computational fluid dynamics}
\newacronym{amr}{AMR}{adaptive mesh refinement}
\newacronym{samr}{SAMR}{block-structured adaptive mesh refinement}
\newacronym{d3q19}{D3Q19}{three-dimensional lattice model with 19 directions (corners missing)}
\newacronym{d3q27}{D3Q27}{three-dimensional lattice model with 27 directions}
\newacronym{mpi}{MPI}{message passing interface}
\newacronym{openmp}{OpenMP}{open multi-processing}
\newacronym{sfc}{SFC}{space filling curve}
\newacronym{simd}{SIMD}{single instruction/multiple data}
\newacronym{cpu}{CPU}{central processing unit}
\newacronym{pdf}{PDF}{particle distribution function}
\newacronym{trt}{TRT}{two-relaxation-time}
\newacronym{id}{ID}{identifier}
\newacronym{aabb}{AABB}{axis-aligned bounding box}

\newcommand{\mpi}{\glsdisp{mpi}{\glsentryshort{mpi}}}

\newcommand{\openmp}{\glsdisp{openmp}{\glsentryshort{openmp}}}
\newcommand{\hpc}{\glsdisp{hpc}{\glsentryshort{hpc}}}

\usepackage[]{todonotes}
\presetkeys%
    {todonotes}%
    {inline}{}

\crefname{section}{section}{sections}
\Crefname{section}{Section}{Sections}

\crefname{subsection}{section}{sections}
\Crefname{subsection}{Section}{Sections}

\crefname{subsubsection}{section}{sections}
\Crefname{subsubsection}{Section}{Sections}

\newcommand{\vv}[1]{\boldsymbol{#1}}

\newcommand{\vphi}{\ensuremath{\vv{\phi}}}
\newcommand{\vc}{\ensuremath{\vv{c}}}

\newcommand{\vmu}{\ensuremath{\vv\mu}}

\newcommand{\phia}{\ensuremath{\phi_\alpha}}

\newcommand{\ha}{\ensuremath{h_\alpha}}

\newcommand{\walberla}{{\textsc{waLBerla}}}

\newcommand{\systemAlAgCu}{\textsf{Al-Ag-Cu}}

\newcommand{\calphad}{{\textsc{Calphad}}}

\newcommand{\skampy}{{\textsc{Skampy}}}

\allowdisplaybreaks

\renewcommand{\cite}[1]{\citep{#1}}

\let\originaleqref=\eqref
\renewcommand{\eqref}{eq.~\originaleqref}

\DeclareRobustCommand{\hl}[1]{#1}
\DeclareRobustCommand{\hlgreen}[1]{#1}

\begin{document}

\title{A Scalable and Extensible Checkpointing Scheme for Massively Parallel Simulations}

\author[1]{Nils Kohl\thanks{nils.kohl@fau.de}}
\author[2]{Johannes H\"otzer}
\author[1]{Florian Schornbaum}
\author[1]{Martin Bauer}
\author[1]{Christian Godenschwager}
\author[1]{Harald K\"ostler}
\author[2,3]{Britta Nestler}
\author[1,4]{Ulrich R\"ude\thanks{ulrich.ruede@fau.de}}
\affil[1]{\footnotesize{Chair for System Simulation, Friedrich-Alexander University Erlangen-N\"urnberg (FAU), Cauerstraße 11, 91058 Erlangen, Germany}}
\affil[2]{\footnotesize{Institute of Materials and Processes, Karlsruhe University of Applied Sciences, Moltkestrasse 30, 76133 Karlsruhe, Germany}}
\affil[3]{\footnotesize{Institute for Applied Materials (IAM), Karlsruhe Institute of Technology (KIT), Straße am Forum 7, 76131 Karlsruhe, Germany}}
\affil[4]{\footnotesize{Parallel Algorithms Project, CERFACS, Toulouse, France}}

\maketitle

\begin{abstract}

Realistic simulations in engineering or in the materials sciences
can consume enormous computing resources and thus require the
use of massively parallel supercomputers.
The probability of a failure increases both with the runtime and with the number of system components. 
For future exascale systems it is therefore considered critical that strategies are developed to
make software resilient against failures. 
In this article, we present a scalable, distributed, diskless, and resilient 
checkpointing scheme that can create and recover snapshots of 
a partitioned simulation domain.
We demonstrate the efficiency and scalability of the checkpoint strategy for simulations with
up to $40$ billion computational cells executing on more than $400$ billion floating point values.
A checkpoint creation is shown to require only a few seconds and
the new checkpointing scheme scales almost perfectly up to \hl{more than $260\,000$ ($2^{18}$)} processes. 
To recover from a diskless checkpoint during runtime,
we realize the recovery algorithms using ULFM MPI.
The checkpointing mechanism is fully integrated in a state-of-the-art high-performance 
multi-physics simulation framework.
We demonstrate the efficiency and robustness of the method with
a realistic phase-field simulation originating in the material sciences \hl{and with a lattice Boltzmann method implementation}.

\end{abstract}

\section{Introduction}\label{sec:introduction}

Modern simulation software must scale to massively parallel computing clusters to provide sufficient memory and computational power. 
Many algorithms that are used to model the respective physical processes take the form of explicit stencil codes that can be executed in parallel. 
Since processor clock rates have ceased to improve, only additional concurrency can be used to
deliver more computational power.
With the increasing number of concurrently working processing units,
the probability of a failure in the system in a certain time period rises as well.
This problem is observed in studies of real world systems \cite{Schroeder2006,Taerat2008,Zheng2011,Martino2014}, examined in theory \cite{Herault2015} and discussed as a grand challenge in the upcoming exascale era of supercomputing \cite{Dongarra2011,Cappello2014}. 

There are different categories of errors that may affect a running system including hardware and software faults or network issues. The error types can additionally be divided into hard errors like hardware outages that cause the affected system or application to crash and soft errors like bitflips that can be introduced by cosmic radiation and may not be noticed by the system but cause wrong results. In the following, we discuss only hard errors with a focus on process faults.

Since modern simulation software applications may run for several hours or even days, as e.g.\ in \cite{Hoetzer2015,SuperMUC16,Steinmetz2016,Bartuschat2015,Bartuschat2014}, 
they are likely exposed to outages. 
Hence, applications that are not prepared to handle these
errors might fail or lead to wrong simulation results.

There are different approaches to develop resilient applications. Two major categories are checkpoint-restart and algorithm-based techniques. Checkpoint-restart methods are based on creating 
regular snapshots of the simulation data that can be recovered after a failure.
The simulation then continues from the recovered backup. 
Consequently, this approach is suited for schemes that yield a sequence of well-defined states.
Hence, it especially fits explicit time stepping schemes like the phase-field method proposed in \cite{Hoetzer2015} or the lattice Boltzmann method (LBM) \cite{Krueger2016}.
Algorithm-based resilience techniques are based on application dependent algorithms
to recompute the data that may have been lost due to a failure.

In this work, we propose a scalable checkpoint-restart method to prepare long running simulation applications with schemes that yield a sequence of well-defined states for hard faults in parallel systems and exemplarily apply them to a large-scale phase-field \hl{and a lattice Boltzmann} simulation application.

\subsection{Outline}

At best, a resilience technique should at the same time be robust, introduce low overhead in terms of memory and runtime and be flexible and easy to implement into existing frameworks. 
To fulfill these requirements, we present and implement a resilient, diskless and distributed checkpointing scheme to regularly create snapshots of the simulation data. 

A checkpoint-restart mechanism with diskless snapshots requires the application to be able to recover from faults during runtime.
Therefore, we use the User Level Failure Mitigation (ULFM) extension to the current \mpi{} standard developed by the \mpi{} Forum that enables applications running with \mpi{} to recover from runtime faults \cite{Bland2012,ULFM2012,ULFM2013}. 
This approach allows us to solely rely on in-memory checkpoints without the need for persistent storage and therefore requires only low overhead to create and restore the snapshots. 

We implement the presented technique in the multi-physics simulation framework \walberla{} \cite{Godenschwager2013}. Using its flexible data structures, we can distribute the snapshots to restrict the memory overhead and to rebalance the workload after process faults to maintain low runtimes.

\subsection{Contribution}

In this work we show that
\begin{enumerate}
\setlength\itemsep{3pt}

\item[(i)] our in-memory checkpointing implementation scales almost perfectly to \hl{$2^{18}$} \mpi{} processes -- \hl{to our knowledge, this is the largest application of this resilience technique to date} (see \cref{fig:weakscalingjuqueen});

\item[(ii)] it only takes a few seconds to create a snapshot of a simulation domain consisting of roughly $40$ billion cells and more than ten times as many floating point values (see \cref{fig:weakscalingsupermuc});

\item[(iii)] we approximately spend less than $4 \%$ of the runtime on checkpoint creation using a theoretically optimal checkpointing frequency on a system with a mean time between failures (MTBF) of one hour (see \cref{fig:wastegraph});

\item[(iv)] the data recovery process scales equally and takes less than a second \hl{(without load balancing)} in all of our test scenarios since it does not involve any communication (see \cref{fig:recoveryemmy});

\item[(v)] we are able to continue a phase-field \hl{and a lattice Boltzmann} simulation from the last checkpoint after process faults during runtime using a ULFM MPI implementation.
\end{enumerate}

The rest of the paper is structured as follows: \Cref{sec:faults} summarizes some related work regarding faults in \hpc{} systems. In \cref{sec:walberla} we will introduce the software design and data structures of the \walberla{} framework which we extend by the presented resilience techniques. \Cref{sec:mpi} covers a proposal for an extension to the current \mpi{} standard that was developed by the \mpi{} Forum and provides the user the means to stabilize the parallel environment after process faults. In \cref{sec:scalablecheckpointrecoveryscheme} we first describe strategies that can be employed to recover an application after a fault in the underlying system. The section then covers the algorithmic details of the checkpointing and recovery schemes that we implemented in the course of this work. In \cref{sec:phasefield} we briefly introduce a real-world application based on the phase-field model that is used to demonstrate and benchmark the covered resilience techniques. Finally, in \cref{sec:benchmarks} we present the results of various benchmarks.
\subsection{Related Work}

\hl{In }\cite{Moody2010}\hl{, the authors propose the Scalable Checkpoint/Restart (SCR) library, that can be integrated into existing parallel applications to extend them with checkpoint-restart mechanisms. While the SCR library supports in-memory checkpoints, it requires the application to start with additional spare processes, that are additionally used when other processes fail. However, the \walberla{} framework does not require a static number of processes since the workload can be efficiently distributed to other processes as described in }\cite{Schornbaum2017}\hl{. Therefore, no resources are wasted in case there are no failures.}

\hl{The Fault Tolerance Interface (FTI) proposed in }\cite{Bautista-Gomez2011}\hl{ can be integrated into existing applications similar to the SCR library. The multi-level checkpointing scheme employed by the FTI is based on file system checkpoints. The FTI library could be used as an extension to the approach presented in this paper since it offers different ways to store the snapshot data. Both libraries can only help to serialize the checkpoints and do not provide a full solution to runtime-resilience combined with adaptive algorithms and load balancing }\cite{Schornbaum2015,Schornbaum2017}\hl{.}

\hl{Diskless and pair-wise checkpointing techniques have been evaluated in }\cite{Chiueh1996,Silvaa1998,Chen2005}\hl{ and more recently in }\cite{Zheng2012}\hl{. The authors of }\cite{Zheng2012}\hl{ propose a quite similar approach to the one presented in this paper. It is based on the Charm++ library and uses a custom MPI extension called Adaptive MPI to achieve runtime resilience. However, since \walberla{} itself provides an ultra-scalable framework for parallel simulations and includes features like load balancing and adaptive mesh refinement (AMR), porting the library to Charm++ would not be straightforward. Apart from that, ULFM already includes mechanisms that decrease the algorithmic complexity of the implementation as it automatically detects process failures as described in }\cref{sec:mpi}\hl{ and handles ongoing communication in the revoked communicator. Combined with already implemented (de-)serialization callbacks that are required for the AMR building blocks of \walberla{}, adding the checkpoint-restart mechanism is only one minor additional step.}
\section{Faults in Parallel Systems}\label{sec:faults}

The increasing complexity and size of modern HPC infrastructure and corresponding software raises questions about the stability of long running applications. 

As we are reaching the exascale era of supercomputing, the frequency 
of faults is expected to rise
\cite{Cappello2014,Dongarra2011}. 
Once the MTBF is in the order of the runtime of a simulation then it becomes likely that a failure strikes. Restarting such applications until successfully completed may incur high cost.

The reliability of actual HPC systems has been studied in e.g.~\cite{Schroeder2006,Taerat2008,Zheng2011,Martino2014}. In \cite{Schroeder2006}, the data from $22$ different HPC systems at the Los Alamos National Laboratory over the course of nine years from $1996$ -- $2005$ is collected. The authors conclude that failure rates in HPC systems seem to be roughly proportional to the number of processors in the systems and rather not depend on the type of the employed hardware. This is still an interesting outcome, although the system sizes are hardly comparable to state-of-the-art clusters. To better understand the link between system outages and job interruptions, \cite{Zheng2011} examines the reliability, availability and serviceability logs as well as the jobs logs that have been collected over $273$ days from a $40$-rack Blue Gene/P system with about $163\,000$ cores in total on more than $40\,000$ nodes. 
The authors observe that jobs with a larger size, i.e.~a larger number of processes
tended to be more affected by system failures than jobs that have smaller size but longer execution times.

In \cite{Herault2015} a theoretical model is presented to show that the failure rate of a parallel system is proportional to the number $N$ of employed nodes. Defining $\mu$ as the MTBF of the system and $\mu_\text{ind}$ as the MTBF of an individual node, the authors show that
\begin{equation}
\mu = \frac{\mu_\text{ind}}{N}
\label{eq:mtbfparallel}
\end{equation}
assuming that $\mu_{\text{ind}}$ is equal on all nodes.

Although the results of the studies are hardly comparable in a quantitative fashion, they show a similar trend and therefore support our assumption: the number of components that is employed when running parallel applications is linked to the MTBF.

\section{Exascale Simulation Software Design}

\label{sec:walberla}

While the strategies we present in this work are generically applicable, the actual implementation 
benefits from a framework that provides sufficient flexibility, extensibility, and appropriate data structures. Likewise, we show that our approach to
resilience is suited to be implemented in large, complex and scalable simulation frameworks.
In particular, we demonstrate that it can be used and is fully operational in current computing practice.
To demonstrate this, we will implement our fault tolerance techniques in the simulation framework 
\walberla{} \cite{Goetz2008,Donath2009,Feichtinger2009,Godenschwager2013,Bartuschat2014,Schornbaum2015,Bauer-python-2016}.

\walberla{} (\textit{widely applicable Lattice Boltzmann from Erlangen}) is a parallel HPC software framework for multi-physics simulations mainly targeting applications based on the LBM. It provides efficient data structures, communication schemes, load balancing and implements performance critical optimizations like SIMD vectorization and intra-node multithreading. 

Besides its support for the LBM to perform fluid simulations and the integrated \textit{PE physics engine} \cite{Goetz2008,Iglberger2009,Preclik2015} which can be used to study particle-laden flows, \walberla{}'s modularity also allows to implement general stencil algorithms on structured grids. 

It is written in C++ and can be employed on all kinds of machines and operating systems as it supports the major compilers and scales from desktops up to large HPC clusters. It uses \openmp{} for shared memory parallelization and \mpi{} for distributed memory communication, while both can be combined in a hybrid approach to run multiple \openmp{} threads per \mpi{} process. A careful performance analysis of \walberla{} on heterogeneous CPU-GPU clusters has been presented in \cite{Feichtinger2015}.

\subsection{Domain Partitioning}
\walberla{} partitions the simulation domain into blocks that may carry arbitrary data including classes that are provided by the framework, C++ standard library data structures or data structures defined by the user.

Each block is assigned to one process. However, a process may own more than one block. Each process is only responsible to execute the simulation on the blocks assigned to it. Consequently, mechanisms like load balancing are realized by distributing the workload in terms of whole blocks.

The structure that holds all the blocks is completely distributed. Thus, every process only carries the data of its assigned blocks and only knows of the blocks that are in their direct neighborhood. No information about the global domain partitioning are stored. This has two implications: the memory allocated on any process only depends on the number of blocks on that process and the data is not stored redundantly in any way, i.e. if block data is lost on one process, it cannot be restored by others. This way \walberla{} is able to scale perfectly with system size, as the memory requirement of each process is independent of the number of processes that run the simulation.

\walberla{} can perform simulations on nonuniform grids \cite{Schornbaum2015}. It provides mechanisms to refine the domain during the setup phase or adaptively during runtime \cite{Schornbaum2017}. Each block can be refined into eight equally sized blocks in parts of the domain that require a higher resolution while a $2$:$1$ balance of the refinement levels of neighboring blocks has to be maintained. It is possible to trigger the refinement during the simulation, for example if it is not known beforehand where the domain will require higher resolution.

Additionally, \walberla{} enables the user to perform runtime load balancing. Different load balancing algorithms can be applied via callback functions. To efficiently handle the data redistribution, the implementation uses a lightweight proxy block data structure without simulation data during the load balancing process. As soon as the load is sufficiently balanced, the actual simulation data is migrated.

\subsection{Communication}

Parallelization and communication in \walberla{} are realized by \openmp{} for shared memory within a node and \mpi{} for distributed memory between nodes. It is also possible to run \walberla{} applications using a hybrid \openmp{} / \mpi{} approach. In this case, the library spawns \openmp{} threads on every \mpi{} process when needed. In order to hide the actual communication interface from the user, an extra communication layer is introduced, that wraps calls to \mpi{}. This layer allows for optimizations - for example if block data were exchanged between two blocks that reside on the same process. In this case \mpi{} functions do not have to be called since memcopy can be used instead. In \walberla{}, data is communicated by packing and unpacking to and from buffers, which are sent and received via \mpi{} respectively. With hybrid \openmp{} / \mpi{}, the data can be packed and unpacked in parallel via multiple \openmp{} threads. This allows for example to create and send packages for every face, edge and corner of a block in different threads.
\section{Resilience in MPI}\label{sec:mpi}

The communication system that is employed by a parallel application plays an important role for handling faults in the underlying environment. It is likely that processes become aware of faults in a parallel system through failing communication. \mpi{} is one of the most common communication systems in parallel applications and is used in \walberla{} for distributed memory communication.

The current \mpi{} specification (version 3.1) does not include mechanisms to deal with failures that are caused by the underlying system \cite{MPIStandard-3.1}. It is rather the implementation that shall be responsible for such failures and resolve them without letting the user notice them at all. Yet, some failures like for example process faults might not be resolvable by the implementation and will therefore result in an error raised by the affected MPI routine.

\begin{table*}
\centering
\footnotesize
\renewcommand{\arraystretch}{1.4}
\begin{tabularx}{0.9\textwidth}{ X | X }
\textit{ULFM Error Codes and Routines} & \textit{Explanation} \\
\hline\hline
\lstinline[]$MPI_ERR_PROC_FAILED$ & Returned when an \mpi{} function cannot operate failure-free due to process failure. \\
\hline
\lstinline[]$MPI_ERR_PROC_FAILED_PENDING$ & Returned when a non-blocking receive function that accepts messages from \lstinline[]$MPI_ANY_SOURCE$ was posted and no send message has matched, while there are potential sending processes that are known to have failed. \\
\hline
\lstinline[]$MPI_ERR_REVOKED$ & Returned if a communication routine is used on a revoked communicator.\\
\hline\hline
\lstinline[]$MPI_Comm_revoke( comm )$ & Marks a communicator as revoked. If a communicator is revoked, all \mpi{} functions that attempt to use it will immediately return the error code \lstinline[]$MPI_ERR_REVOKED$ without performing any communication. Also, all currently running communication routines on the communicator will exit with this error code.\\
\hline
\lstinline[]$MPI_Comm_shrink( comm, newcomm )$ & Creates a new communicator that discards all failed processes. In general, ranks are assigned to different processes on the new communicator.\\
\hline
\end{tabularx}
\vspace{6pt}
\caption{Summary of some of the new \mpi{} error codes and routines introduced by the ULFM extension.}
\label{tab:ulfmroutines}
\end{table*}

The practice of insulating the user from the error handling process is more or less rescinded by an extension to the MPI specification called User Level Failure Mitigation (ULFM) that was recently proposed by the MPI Forum \cite{Bland2012,ULFM2012,ULFM2013}. It targets the recovery from failures like processor faults that result in unresponsive processes.

ULFM offers a lot of flexibility as it does not define a particular recovery mechanism but rather provides the user with routines to bring the MPI environment into a stable state. That means that the user may employ any recovery strategy using the repaired MPI infrastructure including checkpoint-rollback or algorithm-based recovery techniques. The user is also able to reach different levels of recovery. Possibly not all workers in a master-worker environment must be aware of other workers being hit by a fault. In this case, comparably expensive collective failure reporting routines can be avoided while they might be essential in other, homogeneous applications where all processes might for example take part in collective communication.

\Cref{tab:ulfmroutines} shows a summary of the subset of new \mpi{} routines introduced by ULFM that is relevant for our implementation. The recovery pipeline we implement can be divided into three steps:

\begin{enumerate}
\setlength\itemsep{3pt}

\item[(i)] As soon as a process failure is signaled by an \mpi{} routine the communicator is revoked via \lstinline{MPI_Comm_revoke()}. A failure is either signaled 
\begin{itemize}
\setlength\itemsep{3pt}
\item directly (via \lstinline{MPI_ERR_PROC_FAILED} or \lstinline{MPI_ERR_PROC_FAILED_PENDING}), if a process tries to communicate with a failed process or
\item indirectly (via \lstinline{MPI_ERR_REVOKED}), if a process does not directly communicate with a failed process, but another process does, and revokes the communicator.
\end{itemize}
This step is necessary since we need to inform all processes that a fault has happened.

\item[(ii)] The communicator is shrunk via \lstinline{MPI_Comm_shrink()} in order to discard all failed processes from the old communicator. The new communicator is no longer revoked.

\item[(iii)] An application-level recovery mechanism is started. In our case, the last checkpoint is recovered.

\end{enumerate}

Although the extension is a proposal and not yet part of the actual standard, an implementation of a subset of the proposed concepts already exists in form of a fork of the OpenMPI implementation. This allows for testing and preliminary integration into production software like \walberla{}.

\section{A Scalable Checkpoint-Recovery Scheme}\label{sec:scalablecheckpointrecoveryscheme}

In this section we will introduce the checkpoint-rollback recovery strategy as implemented in \walberla{}. We will discuss the general approaches, advantages and disadvantages of checkpoint-rollback recovery and also briefly introduce algorithm-based recovery techniques as an alternative to the chosen approach. Then in the second part of this section we will describe the details of our implementation.

\subsection{Recovery Strategies}\label{sec:recoverystrategies}

If a fault in a parallel system affects a component that is simultaneously used by an application it might lead to lost data, errors or crashes during the program's runtime. Especially for long running simulations a restart can be very costly. However, regarding massively parallel software it is likely that only a fraction of the used components is actually affected by the failure. Thus, respective applications might recover by dismissing or replacing the erroneous components and recovering potentially lost data from backup or by recomputing.

In this subsection, we aim to present some of the techniques and algorithms that can be used to restore lost data and to continue the execution despite failures in the system. First of all, we will examine checkpoint-rollback recovery. Finally, we will give a brief overview of algorithm-based recovery techniques. 


The most widely used and most straightforward techniques to recover from process failures in HPC applications are based on restorable checkpoints of an application state \cite{Di2014,Moody2010,Bautista-Gomez2011,Zheng2012,Herault2015,Elnozahy2002,Young1974,Daly2006,Chandy1975,Treaster2005}. When a failure happens, the application rolls back to the last valid checkpoint and recomputes the data that was lost due to the failure and the rollback. The major advantages of this technique are its simplicity and that it can be used as a general-purpose method as it is not tailored to a specific algorithm. The latter point is a benefit of the checkpoint-rollback technique introduced in this work. This way, fault tolerance is not limited to certain algorithms, but available for all applications that use the framework's data structures. In our case this includes LBM algorithms as well as the phase-field method that will be employed as an example application later in this work.

The idea of checkpoint-rollback recovery can be realized in many different ways depending on the underlying application. Some (parallel) applications might require a \textit{coordinated} checkpointing algorithm, meaning that the processes must create a snapshot of the application's state at the same time in order to assure a certain consistency between all processes. For other applications, an \textit{uncoordinated} scheme, where all processes create checkpoints but not necessarily at the same time might be sufficient \cite{Herault2015}. The latter technique has the advantage that the processes do not have to spend time to coordinate the checkpoint and that potentially shared I/O resources are not simultaneously used by all of them. As a tradeoff, hierarchical strategies might be employed like for instance checkpointing in groups of processes.

As orthogonal criteria, the data to be stored, the type of storage and the level of redundancy may differ. Some strategies might consider dumping the entire application memory (\textit{system-level checkpoints}) \cite{Hargrove2006} while others only store particular data structures chosen at application level (\textit{application-level checkpoints}) \cite{Moody2010,Bautista-Gomez2011,Zheng2012}, which might save resources in terms of storage space and checkpointing duration. On the other hand, a system-level checkpointing mechanism might decrease the complexity of the application itself as external tools can be used more easily.

While it might be beneficial to store the snapshot to \textit{disk} in order to restart the whole application at a later point in time or to inspect the data offline, time can be saved by only backing up the data in the \textit{main memory} of different processes, provided the system can tolerate failures without halting the application \cite{Zheng2012}. \hl{Both approaches can be combined resulting in a multi-level checkpointing scheme as presented in} \cite{Moody2010,Bautista-Gomez2011}. For some techniques, tradeoffs between \textit{redundancy} of the data can be made by storing parts of it on only some processes. This also poses a tradeoff between resource savings and resilience of the system as only some processes carry a certain part of the domain and a fault will result in ultimate loss of data if all processes of that group fail.

In this article we implement a version of a \textit{coordinated} \textit{application-level} scheme with \textit{in-memory checkpoints}. As the cell updates in stencil based algorithms depend on the cells' neighbors, the checkpoints must be created at the same time during the simulation (coordinated). Only this way we can ensure that the simulation \hl{is kept in a consistent state} after a rollback. To keep the memory footprint low, only data structures, that cannot be recreated automatically from other snapshot data are stored during a checkpoint. The implementation allows for flexible redundancy schemes by offering the user the option to register callback functions with custom rules defining the number of snapshot copies and the processes they shall be stored on. \Cref{sec:implementation} contains more details on the implementation of the checkpointing scheme.


One alternative to checkpoint-rollback recovery  
are so called {\em algorithm-based} fault tolerance (ABFT) techniques. They are not based on 
checkpoints of the domain but aim to recover lost data 
using application specific algorithms.  
This has the advantage that in general little to no time is 
spent during failure-free periods and therefore the computation time without faults does not increase significantly. 
However, as ABFT depends on the application, the recovery algorithms 
must be designed carefully and are in general not 
usable on other applications. 
In many cases, no ABFT algorithms are known.

A successful example of ABFT 
in numerical linear algebra 
is the recovery of matrix entries via checksums. By storing additional vectors that contain the checksums of a matrix' rows and columns, lost or corrupt entries can often be reconstructed \cite{Huang1984,Bosilca2008,Yao2011}. 

\hl{Another successful ABFT approach 
to recover data loss in parallel multigrid algorithms is presented in }\cite{Huber2016}\hl{ 
for the Laplace equation but extends to more general problems.
Just as the checkpoint-rollback algorithm it only assumes that the domain is partitioned in blocks,
whose data may be lost when a process fails.
However, different from a checkpointing algorithm, the ABFT technique is based on
a fast recomputation of lost data. 
The specific technique relies on only the ghost layer data being stored redundantly so that
the lost data can be recomputed by a {\em local} multigrid algorithm.}

In the phase-field simulations as described in \cite{Bauer2015,hoetzer15c} the various fields can be recovered from partial checkpoint data or possibly algorithm-based by exploiting physical correlations.
Depending on the field and the available checkpoint data, the fields can be recovered without information loss, otherwise the state of the fields is calculated in an approximated way using e.g.\ analytic equations.
For example in the phase-field model used here, one of the concentration fields can be recovered without
loss of information if the other concentration fields are available by exploiting mass conservation.
In the case that the concentration fields are not available,
the field state can be calculated using the phase-fields and the Gibbs energies in an approximate fashion.
Similarly, the phase-fields can be recovered from partial checkpoint data.
Also the temperature fields can be recovered in an approximate fashion if the checkpoint does not exist by using analytic profiles.

\subsection{Implementation}\label{sec:implementation}

We implement a diskless and coordinated application-level checkpointing strategy to achieve a flexible, robust and scalable solution regarding faults in the underlying system. The method builds on the ability to tolerate faults during runtime. We reach this by using the prototype ULFM \mpi{} implementation that was introduced in \cref{sec:mpi}. In this section, we will present the architecture and algorithms of our specific approach. 

\subsubsection{Checkpoints.}

\paragraph{Diskless Checkpointing.} The chosen strategy employs in-memory checkpoints. This means that the data is backed up by distributing it to the main memory of the processes. Obviously, in-memory checkpoints lead to data loss as soon as a process that carries the respective memory fails or when the application terminates. Therefore, it is crucial that a respective implementation is capable of recovering from a fault during runtime. If the whole application terminated, no backup would be left to be restored. However, one could for instance additionally implement checkpointing to disk, \hl{for example using techniques presented in} \cite{Moody2010,Bautista-Gomez2011,Hargrove2006} at a lower frequency to protect the simulation against failures that strike the whole system. 

The main advantage of in-memory checkpoints is that they are not limited to possibly slow I/O of disk storage and are not affected by other nodes in the cluster using the same device to write and read data at the same time. This way it is possible to maintain relatively low overhead.

\paragraph{Custom Data Structures.} The checkpointing frequency and execution is maintained by a callback, which is automatically invoked with a parametrized period between two iterations. During the setup, this checkpointing callback accepts callbacks for every entity that needs to be backed up. \hl{Here, the term entity stands for a class or data structure, the instances of which are serializable. Examples for entities that need to be backed up are the blocks of the domain, including their simulation data, the respective mesh data as well as metadata like information about neighboring blocks. Others are for instance timers that need to be reset to the timestamp of the last valid checkpoint. Entities that shall be backed up need to be registered at the checkpointing callback.}

All entities that shall be restorable after a fault, must provide three callback functions: one that creates a snapshot, one that restores the snapshot after a fault and one that swaps the snapshot buffers. The latter one will be explained later in this section. In this way, each entity is responsible for the snapshot creation of its own data. This has three advantages: there is no need for a class or function that can snapshot arbitrary data, the snapshots are highly customizable for every entity and their data stays decoupled from the checkpointing mechanism. On the other hand, this means that the three callback functions need to be implemented individually for different entities. 

\paragraph{Redundancy.} As stated in previous sections, the domain partitioning that is implemented in \walberla{} results in fully distributed data structures that evidently lie in main memory. Opposed to disk storage, main memory is in general lost as soon as the respective process or node faults. In order to securely store the data, we therefore need to employ a certain level of redundancy. When one process fails, its data must be backed up somewhere else. For in-memory snapshots, the data must be stored on another process that is still alive. Otherwise the backup is entirely lost and the application cannot recover from the failure. However, for large scale simulations it is in general not possible to store the backup in an all-to-all fashion, meaning that every process stores a backup of the data of all other processes. A lower level of redundancy must be implemented - only some copies of the backups from one process are made and stored in the memory of some other processes \hl{as presented in} \cite{Zheng2012,Chen2005,Chiueh1996,Silvaa1998}. This unfortunately leads to a certain tradeoff regarding resilience of the system: it cannot always be guaranteed that there is a backup available in the part of the system that survived a fault. In other words, if all processes that carry the backup of a failed process fail as well, that data is lost. Indeed, with the growing number of nodes in modern parallel systems, the probability for data loss shrinks, if the data is well distributed. The risk can for instance be minimized by storing backups on different nodes that are rather unlikely to fail at the same time. 

\begin{figure*}[t]
\centering
\resizebox {\textwidth} {!} {
\includegraphics[width=1.0\columnwidth]{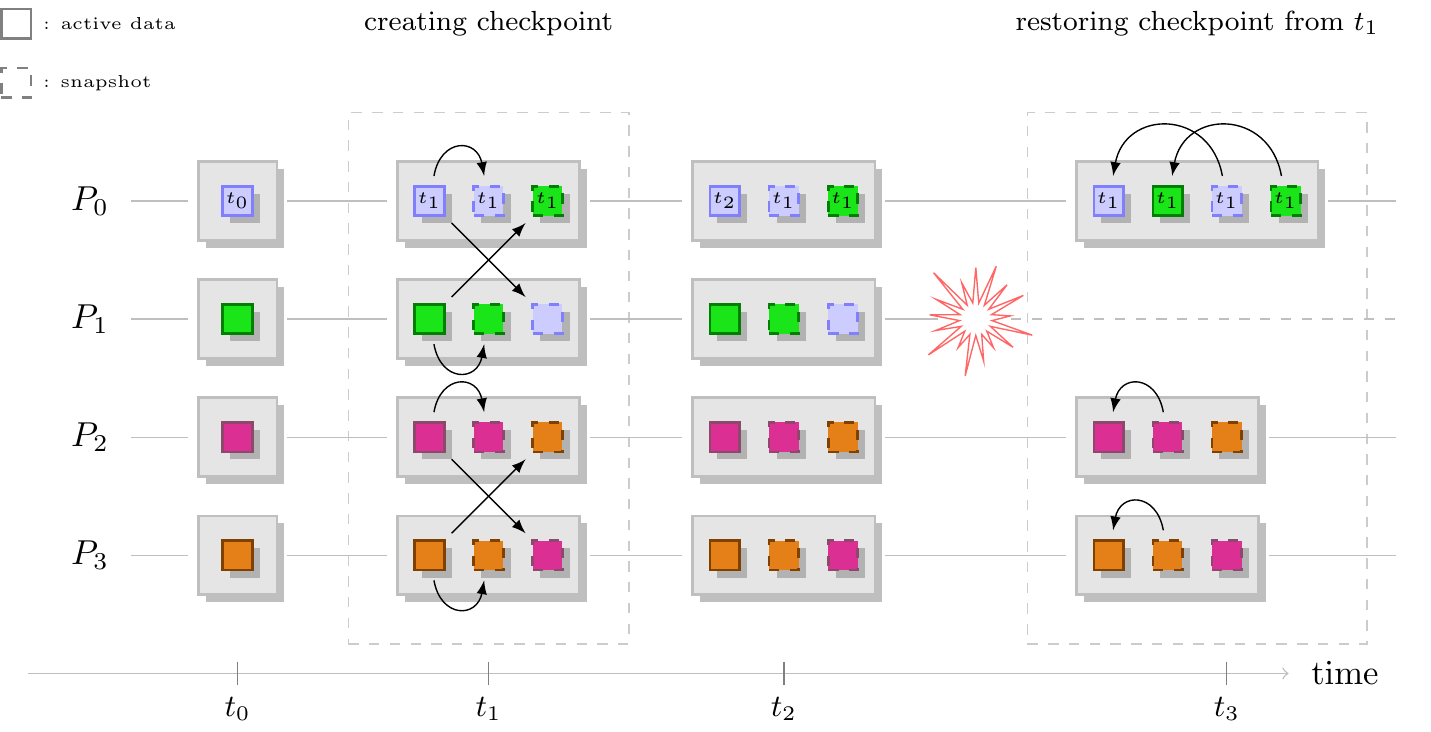}
}
\caption[Illustration of a pair-wise checkpointing strategy.]{Illustration of a pair-wise checkpointing strategy. At time step $t_1$ each process creates a snapshot of its part of the domain and exchanges a snapshot with its partner process. The processes now carry a distributed checkpoint of the whole domain. After the fault, the living processes restore the checkpoint from $t_1$. This is possible without inter-process communication since each process stored a snapshot of its own domain. $P_0$ is now responsible for the data that was processed by $P_1$ before the fault. \hl{Typically, a load balancing step is required right after the snapshot is recovered to maintain maximal simulation performance (see }\cref{sec:loadbalancing}).}
\label{fig:snapshotstrategy}
\end{figure*}

To decouple the distribution scheme from the actual checkpointing and recovery routines and to offer the user the flexibility to implement custom distribution algorithms, the respective functions that determine the distribution rules can be registered as callbacks. The user needs to choose a scheme that determines the ranks to which the current process shall send its copies and the ranks it will receive copies from. An example for a distribution callback that stores one copy of the backup at the `opposing' rank (shifted by $N_\text{proc} / 2$, \hl{similar to} \cite{Zheng2012}) can be seen in \cref{algo:snapshotpairstrategycreation}. Since nodes typically carry consecutive \mpi{} ranks, this method guards against single-node failures.

To improve the performance, each process stores one copy of its own checkpoint. This way, less data has to be exchanged during the checkpointing step and no data has to be exchanged during the recovery process \hlgreen{(without load balancing)} as can be seen in \cref{fig:snapshotstrategy}.

\begin{algorithm}[t]
\begin{spacing}{0.8}
\caption[Program flow of a pair-wise snapshot distribution]{\footnotesize\itshape
Program flow of a pair-wise snapshot distribution. The function determines the rank the current process shall send its data to and the rank it receives its data from, so that the respective send and receive schedules can be invoked.
}
\label{algo:snapshotpairstrategycreation}
\end{spacing}
\footnotesize
\DontPrintSemicolon
\vspace{3pt}
\SetKwFunction{SD}{PairwiseSnapshotDistribution}
\SetKwProg{Fn}{Function}{}{end}
\SetKw{Out}{out}
\Fn{PairwiseSnapshotDistribution}
{
determine the number of processes $N_{\text{proc}}$

determine the rank of the current process

\If{$N_{\text{proc}} > 1$}
{
    $\text{shift} = N_{\text{proc}} / 2$

    $\text{sendTo} = (\text{rank} + \text{shift}) \mod N_{\text{proc}}$

    \uIf{$\text{shift} > \text{rank}$}
    {
        $\text{recvFrom} = N_{\text{proc}} - (\text{shift} - \text{rank})$
    }
    \Else
    {
        $\text{recvFrom} = \text{rank} - \text{shift}$
    }
}
\Return (sendTo, recvFrom)
}
\end{algorithm}

\paragraph{Resilient Checkpointing.} When the checkpointing callback functor is invoked to create a snapshot, it calls all registered callbacks that are needed to create the checkpoint of the respective entities like blocks and timers. This way, one after another, all instances that have been attached via a respective callback create snapshots of their data. 

In order to make sure that the snapshots of all data structures are consistent on all processes, the mechanism must be robust to faults during the checkpointing process. If for instance a snapshot of the domain has been successfully created while the snapshot of a timer that is used to influence time dependent properties of the simulation was not, the simulation might not behave as expected when the application rolls back to that data. 

To solve this problem, we implement a double-buffer model for the snapshot creation and recovery. Two snapshot buffers instead of one are maintained by each entity that requires data exchange between processes to create a redundant backup of its data (no exchange is needed for instance if the entity's data is equal on all processes in the same iteration). One of those buffers serves as a read-only container, carrying the snapshot that shall be restored upon a fault. This buffer must not be touched during the snapshot creation. The other one is writable and shall be used when a new checkpoint is created. After a snapshot has been created in the writable buffer, all processes perform a handshake to check if any process faults occurred before or during the checkpoint creation. If there was a fault, the read-only buffer still carries a valid snapshot from the last checkpoint which is then recovered. If there were no faults and all processes succeeded creating the checkpoint, the buffers can be swapped to update the read-only buffer with the new checkpoint. Here, the already mentioned callback function that shall swap these buffers comes into play: its purpose is to swap the buffers (preferably by swapping the pointers accessing them to avoid copying data) so that the former read-only buffer becomes the writable one that will be used the next time a snapshot is created and the writable buffer becomes read-only. Since there is no communication involved, the buffer swapping procedure cannot be affected or interrupted by faults in the system. The full resilient checkpointing algorithm is shown in \cref{algo:failureproofcheckpoint}.

\begin{algorithm}[t]
\begin{spacing}{0.8}
\caption[Program flow of the resilient checkpointing procedure]{\footnotesize\itshape
Program flow of the resilient checkpointing procedure using the double-buffer model.
}
\label{algo:failureproofcheckpoint}
\end{spacing}
\footnotesize
\DontPrintSemicolon
\vspace{3pt}
\SetKwFunction{HS}{Handshake}
\SetKwFunction{RLC}{RecoverLastCheckpoint}
\SetKwProg{Fn}{Function}{}{end}
\Fn{CreateResilientCheckpoint}
{
\ForAll{registered entities}
{
    create checkpoint in writable buffer \label{algo:snapshotcreation}
}
\tcp{
The following handshake has two purposes:\\
- it assures that all processes\\ 
~~finished checkpointing\\
- it is used to inform all processes\\
~~of potential faults in the system.
}
call \HS \label{algo:handshake}

\uIf{process(es) failed}
{
    \tcp{Since no buffers have been swapped yet, \\
    the read-only buffer still carries \\
    the previous snapshot.}
    call \RLC \label{algo:recovery}
}
\Else
{
    \tcp{A fault cannot prevent the buffers\\
    from being swapped since no\\
    communication is necessary here.}
    \ForAll{registered entities}
    {
        swap checkpoint buffers \label{algo:swapbuffers}
    }
}
}
\end{algorithm}

\subsubsection{Fault Propagation and Recovery.}
\label{sec:recoveryimplementation}

With a working checkpointing implementation in place, we will now cover what happens when a fault actually strikes. The recovery can be divided into two steps: the recovery of the MPI environment and the recovery of the lost data.

\hlgreen{Failures are detected by the \mpi{} library and signaled to \walberla{} via errors that are raised by ULFM MPI routines. After a failure happens, a process will notice it as soon as it tries to communicate with a dead process via an MPI function. In this sense, the integration of fault detection into the \walberla{} framework does not require the simulation framework itself to actively ping the parallel environment to test if processes died.} \hl{Therefore, the existing communication complexity is not affected by the integration of fault detection mechanisms.}

Although almost all \mpi{} communication routines are wrapped by respective \walberla{} library functions, \mpi{} routines could still be called directly from an application. Therefore, if errors were checked via return codes, some of them might bypass the \walberla{} library. Also, it is very cumbersome to handle return codes throughout a large code base or application. Fortunately, the MPI standard provides the routine \lstinline{MPI_Comm_set_errhandler()} to register a callback function to a communicator that is called when an MPI function on that communicator raises an error. \hlgreen{This way it is possible to handle MPI errors} \hl{that signal process failures} \hlgreen{in one place.}

After a fault has happened, we must first revoke the affected communicators to inform all processes of the fault and secondly we need to shrink them to stabilize the MPI environment. A straightforward approach would be to simply call \lstinline{MPI_Comm_revoke()} and \lstinline{MPI_Comm_shrink()} and restore the snapshots in the registered callback function and return from the callback in order to continue the application. However, this has a severe drawback: as it is not predictable when, respectively where in the code a fault strikes, different processes might resume the application at different places and with different states which most likely results in chaotic behavior.  

Therefore, we use a different approach that ensures a deterministic recovery process. Instead of triggering the recovery mechanism directly, the callback only throws an exception. This exception is then caught in the main program loop as shown in \cref{algo:trycatch}.

\begin{algorithm}[t]
\begin{spacing}{0.8}
\caption{\footnotesize\itshape
Recovery mechanism via exception handling in the main program loop. The checkpoint recovery also adapts the current iteration step. \hl{The try-catch approach also functions as a interface to allow to react to process faults at application level if needed, for instance to calculate the simulation time that was lost due to the failure.}
}
\label{algo:trycatch}
\end{spacing}
\footnotesize
\DontPrintSemicolon
%
\SetKwProg{Fn}{Function}{}{end}
\SetKwBlock{Try}{try}{}
\SetKwProg{Catch}{catch}{}{end}
\Fn{RecoveryMechanism}
{
\While{current step $<$ number of steps}
{
    \Try
    {
        single step 
    }
    \Catch{ProcessFaultException}
    {
        stabilize parallel environment

        recover last checkpoint
    }
}
}
\end{algorithm}

There are two main steps to be performed in the catch block. First, the parallel environment must be brought back into a consistent state. We achieve this by calling the respective ULFM MPI routines (\lstinline{MPI_Comm_revoke()} and \lstinline{MPI_Comm_shrink()}). During this process, information about the fault can be collected. In our case, this includes information of the rank reassignment that may happen during \lstinline{MPI_Comm_shrink()}. Both, the information on the fault as well as the method that was used to distribute the checkpoint data are used to determine how the checkpoint shall be restored. An algorithm that calculates the correct rank assignment for the recovery is shown in \cref{algo:snapshotrestore}. 

\begin{algorithm*}[t]
\begin{spacing}{0.8}
\caption[Program flow of a pair-wise snapshot recovery]{\footnotesize\itshape
Program flow of a pair-wise snapshot recovery distribution. Every backed up block on a process also contains its origin in form of a rank. If this rank is passed to the shown function, the function returns the rank, the data shall be restored on. The result is equal on all processes. Therefore a process can plug in the origins of its backed up blocks and compare the result to its own rank. If it is equal, that process needs to restore the respective block and add it to its own.
}
\label{algo:snapshotrestore}
\end{spacing}
\vspace{3pt}
\footnotesize
\DontPrintSemicolon
\SetKwFunction{SD}{PairwiseSnapshotRecovery}
\SetKwFunction{AB}{Abort}
\SetKwProg{Fn}{Function}{}{end}
\tcp{%
$N_{\text{proc}}^{t-1}$: number of processes before the fault\\
$R^{t-1}$: a rank of a process that existed before the fault\\
$P(R^{t-1})$: the process with rank $R^{t-1}$ before the fault\\
$R_\text{reassignment}(\cdot)$: a function that returns a process' new rank,\\
\phantom{$R_\text{reassignment}(\cdot)$:} given its rank before the fault
}
\Fn{\SD{$R^{t-1}$, $R_\text{reassignment}(\cdot)$}}
{
\tcp{%
Ranks on surviving processes do not necessarily equal the ranks before faults.\\
Therefore, the process that was identified via rank $R^{t-1}$ before the fault could either\\
- still carry rank $R^{t-1}$ ($R^t = R^{t-1}$),\\
- carry a different rank ($R^t \neq R^{t-1}$)\\
- or not exist anymore.
}
\uIf{$P(R^{t-1})$ has not survived}
{
    \tcp{
    We need to find the rank $R_\text{backup}^{t-1}$ that received the checkpoint data from $R^{t-1}$.\\
    For the pairwise checkpointing scheme as shown in \cref{algo:snapshotpairstrategycreation}\\
    this works as follows:
    }
    shift = $N_{\text{proc}}^{t-1} / 2$

    $R_\text{backup}^{t-1} = (R^{t-1} + \text{shift}) \mod N_{\text{proc}}^{t-1}$

    \uIf{$P(R_\text{backup}^{t-1})$ has not survived}
    {
        \tcp{Checkpoint not restorable as only one copy was made.}
        call \AB
    }
    \Else
    {
        \tcp{Determine the new rank $R_\text{backup}^t$ of $P(R_\text{backup}^{t-1})$.}
        $R_\text{backup}^t = R_\text{reassignment}(R_\text{backup}^{t-1})$

        \Return{$R_\text{backup}^t$}
    }
}
\Else
{
    \tcp{Determine the new rank $R^t$ of $P(R^{t-1})$.}
    $R^t = R_\text{reassignment}(R^{t-1})$

    \Return{$R^t$}
}
}
\end{algorithm*}

Then all snapshots can be restored. This typically also involves setting the current iteration step to the value it had when the snapshot was created. That way, the main program loop can be continued from the restored checkpoint. \hl{Additionally, the catch block (see }\cref{algo:trycatch}\hl{) allows for straightforward access to the recovery process at application level.}

\subsubsection{Memory Usage.}

\hl{The described approach increases the memory consumption of the respective application. It is affected by two factors. Depending on the redundancy scheme, additional copies of the simulation data need to be stored in main memory. For the pair-wise checkpointing scheme as discussed above, assuming that each process carries the same amount of data (i.e. the domains of all processes are of equal size), the memory consumption triples: each process needs to store the actively used domain, a snapshot of its own and one of its partner's domain.}

\hl{If the resilient checkpointing algorithm using the double-buffer model is implemented, each snapshot creates a footprint of twice its size. In this case, the overall memory usage of the application is increased by a factor of five when using the pair-wise checkpointing scheme with resilient checkpointing as compared to no checkpointing.}

In general, the per process memory usage of the scheme calculates as
\begin{equation}
\text{\textsc{Mem}} = S(1 + 2R)
\end{equation} 
where $S$ is the memory consumption of the local domain (assuming it is equal on all processes) and $R$ the number of copies ($R=2$ for the pair-wise scheme). Therefore it rises linearly with the redundancy.

\hl{However, considering that due to the diskless implementation the time for the checkpoint creation is expected to be relatively small, one could go without the resilient checkpoint creation to lower the memory consumption to $\text{\textsc{Mem}} = S(1 + R)$ i.e. a factor of three (using the pair-wise scheme) opposed to the memory consumption without checkpointing. In this case, the application cannot be resumed if a fault strikes during the checkpoint creation.

In typical long-running LBM or phase-field applications the memory capacity of each node is not fully exhausted. In such explicit time-stepping schemes, high update frequencies can be achieved by increasing the performance on single processes. In this sense the applications are often run close to the strong scaling limit, i.e. with a comparably small subdomain per process. Frameworks like \walberla{} are highly optimized for such scenarios and will run with good efficiency even when each processor operates only on small subdomains. Therefore the memory increase of the checkpointing method is tolerable for an important class of applications.}

\subsubsection{Load Balancing.}\label{sec:loadbalancing}
\hl{Since the framework's block structure is independent from the underlying process distribution, blocks can be migrated among processes without affecting the simulation results. Thus, there is no need to continue a simulation with the same amount of processes that it ran on before a fault.}

\hlgreen{However, with the pair-wise checkpointing approach, after the recovery, the workload of a failed process is shifted to the surviving process that restored the respective snapshot of the dead process' data. Thus, we can expect a load imbalance right after the recovery process. We therefore need to introduce an additional step to balance the workload among the surviving processes. }

\hl{In large scale computing, i.e. when resilience may become relevant, the loss of a single node will reduce the system performance marginally provided the load can be redistributed.}

Efficient and scalable load balancing techniques are already presented in \cite{Schornbaum2017} and implemented in the \walberla{} framework.

\subsubsection{Checkpointing Frequency.}

An essential parameter that needs to be adapted regardless of the chosen scheme is the frequency of the checkpoints. In \cite{Young1974,Daly2006,Herault2015} the authors derived different but similar approximations to the optimal checkpointing frequency $f_\text{\textsc{opt}}$.

A first-order approximation to the optimal checkpointing interval $T_{\text{\textsc{fo}}}$, i.e. the time between two checkpoints is derived in \cite{Herault2015}. The goal is to minimize the time that is spent checkpointing as well as the time that it takes to recompute values after a rollback to the last valid checkpoint. The resulting first-order approximation is given by
\begin{equation}
\frac{1}{f_\text{\textsc{opt}}} = T_{\text{\textsc{fo}}} \approx \sqrt{2 \mu C}
\label{eq:optimalcheckpointingfrequency}
\end{equation}
where $\mu$ is the MTBF of the underlying system and $C$ the time it takes to create a valid checkpoint of the domain. The model fits applications that perform a coordinated checkpointing scheme. It is important to note that this approximation is only valid if the MTBF $\mu$ is large compared to the time spent checkpointing ($C$). 

The approximation has to be used carefully as it gives only a rough estimate for a theoretical system and relies on some assumptions including a constant MTBF $\mu$ and checkpointing duration $C$ during the life time of an application. Also, it might be hard to estimate $\mu$ for a specific target system. In this sense the estimate \eqref{eq:optimalcheckpointingfrequency} may only serve as an orientation to experiment with different checkpointing intervals.

\section{Phase-Field Method as an application to real-world computational materials design}\label{sec:phasefield}
To demonstrate the practicability of the proposed strategies, we employ them to a phase-field simulation as discussed in detail in \cite{Hoetzer2015}. In this section we briefly summarize the phase-field method.

The properties of materials are related to their microstructure.
For the development of tailored materials, it is of high interest to understand the fundamental mechanisms controlling the microstructure evolution.
To investigate the mechanisms under the influence of different physical effects, the phase-field method has been established as powerful tool \cite{hoetzer2016}. 

\begin{figure}[t]
 \centering
 \includegraphics[width=1.0\columnwidth]{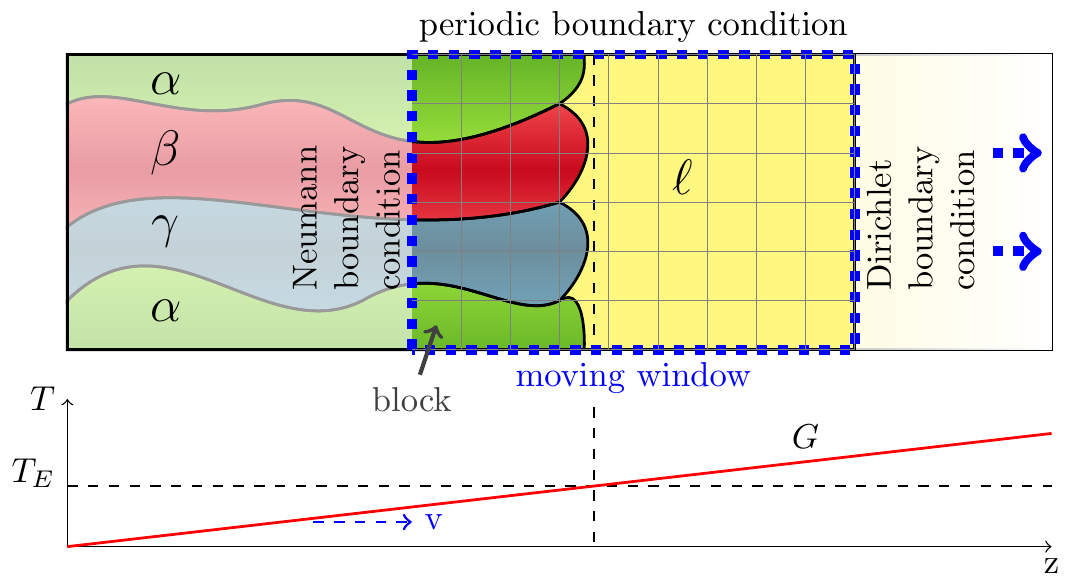}
 \caption{Setting to simulate the ternary eutectic directional solidification based on \cite{hoetzer2017}. Thereby, the melt $\ell$, consisting of three components, solidifies in the three phases $\alpha$, $\beta$ and $\gamma$. In dashed blue \hl{the moving window technique} \cite{Vondrous2014} with the block-structured grid is highlighted. Below, in red the moving analytic temperature gradient is shown. }
\label{fig:setting}
\end{figure}

During the directional solidification of ternary eutectic alloys, various patterns consisting of three solid phases $\alpha$, $\beta$ and $\gamma$ evolve from the melt $\ell$  \cite{Dennstedt2012}.
Due to the wide range of forming patterns in the microstructure, ternary eutectic alloys are  interesting for technical products and scientific investigations \cite{hoetzer2016,Kurz1975}.
The setup for the simulation of directional solidification is schematically depicted in \cref{fig:setting}. 
By applying a moving temperature field with a constant gradient, the solidification direction and velocity are controlled.

\begin{figure}[t]
 \centering
 \includegraphics[width=1.0\columnwidth]{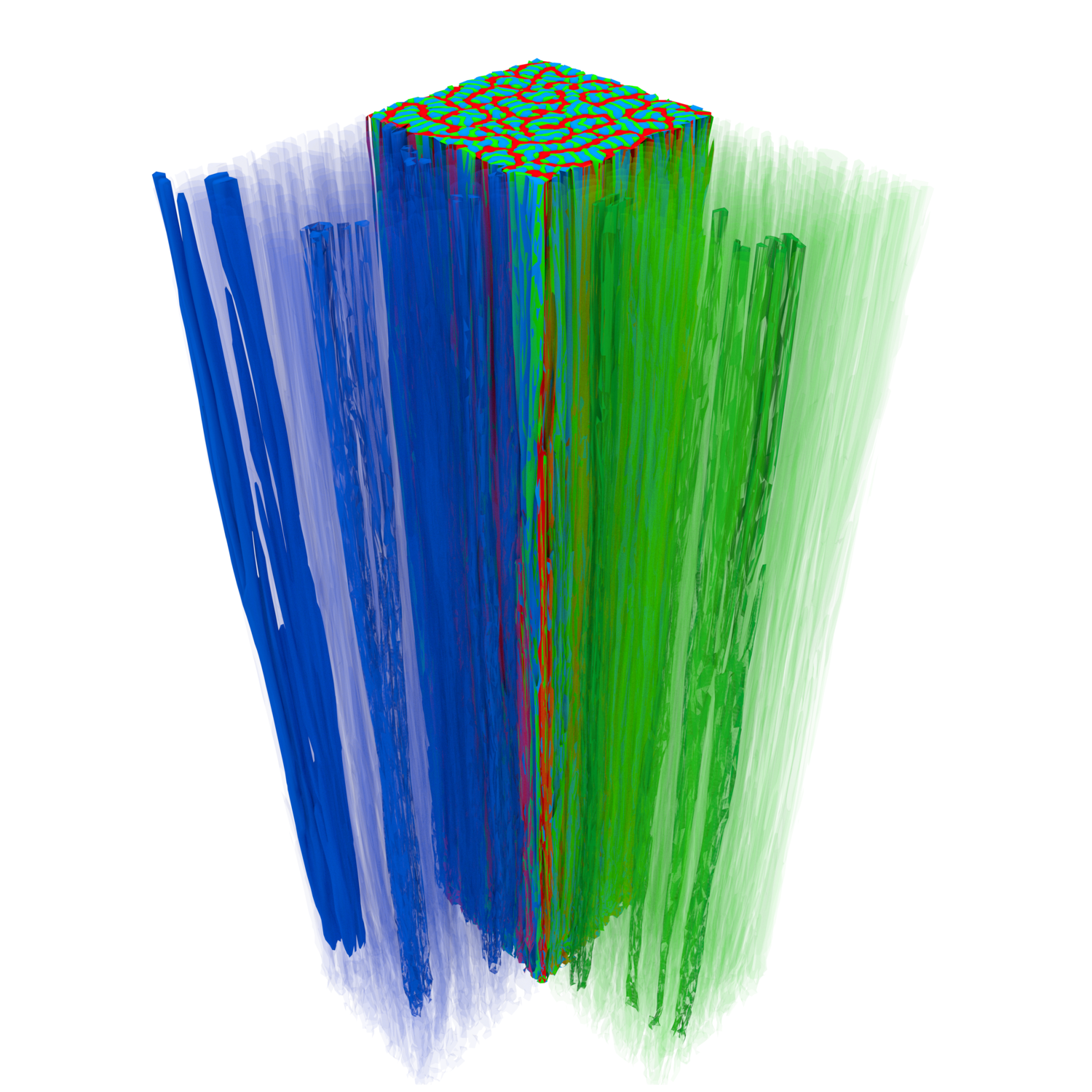}
 \caption{Phase-field simulation of the direction solidification of the ternary eutectic system \systemAlAgCu{} in a $12\,000\times12\,000\times65\,142$ voxel cell domain which was calculated with $19\,200$ cores on the SuperMUC system.
 A detailed discussion is presented in \cite{Hoetzer2015,hoetzer17}.}
\label{fig:pf-sim}
\end{figure}

To investigate the directional solidification of ternary eutectic alloys, the phase-field model based on the Grand potential approach from \cite{Choudhury2012_2} is used.
The evolution equations for the $N=4$ phase-fields $\nicefrac{\partial\phia}{\partial t}$, for the  $K=3$ chemical potentials $\nicefrac{\partial\mu}{\partial t}$ and for the analytic moved temperature gradient  $\nicefrac{\partial T}{\partial t}$ are given as: 
\begin{align}
\tau\epsilon \diffp{\phia}{t} = & 
\underbrace{
-\epsilon T \left(\diffp{a(\vv\phi,\nabla\vv\phi)}{\phia} -  \nabla \cdot \frac{\partial a(\vv\phi,\nabla \vv\phi)}{\partial\nabla\phia} \right) 
}_{:=r_\alpha} 
\nonumber \\
& 
\underbrace{ 
- \frac{T}{\epsilon} \diffp{\omega(\vv\phi)}{\phia}
- \diffp{\psi(\vv\phi,\vv\mu,T) }{\phia}
}_{:=\Gamma_\alpha}  
-\frac{1}{N}\sum^N_{\beta=1} \left( r_\beta + \Gamma_\beta \right) \label{eq:phi_evolution}, \\
\diffp{\vmu}{t} = & 
\left[ \sum_{\alpha=1}^N \ha(\vphi) \left(\frac{\partial \vc_\alpha(\vmu,T)}{\partial \vmu} \right)\right]^{-1} 
\nonumber \\
&
\Biggl( \nabla \cdot \Big(\vv{M}(\vphi,\vmu,T)\nabla {\vmu} - \vv{J}_{at}(\vphi,\vmu,T) \Big) \nonumber\\
& 
- 
\sum_{\alpha=1}^N \vc_\alpha (\vmu,T) \frac{\partial \ha(\vv\phi)}{\partial t} 
\nonumber\\
&
-
\sum_{\alpha=1}^N 
\ha(\vphi) \left(\frac{\partial \vc_\alpha(\vmu, T)}{\partial T} \right) \diffp{T}{t} \Biggr),
\label{eq:chem_pot_evolution}\\
\diffp{T}{t} =&  \diffp{}{t} \left(T_0 + G(z - vt)\right) = -Gv\label{eq:t_evolution}.
\end{align}
The evolution equation for the chemical potentials \eqref{eq:chem_pot_evolution} alone results in $1\,384$ floating point operations per cell and $680$ Bytes that need to be transferred from main memory \cite{Bauer2015}.
Further details of the phase-field model are presented in \cite{Hoetzer2015,Choudhury2012_2}.
The discretizations in space with a finite difference scheme and in time with an explicit Euler scheme are specified in \cite{hoetzer15-3}.
To efficiently solve the evolution equations on current HPC systems, the model is optimized and parallelized as proposed in \cite{Bauer2015}.
Besides explicit vectorization and parallelization with \mpi{}, \hl{a moving window approach} \cite{Vondrous2014} is implemented on top of the block structured grid data structures of \walberla{} as depicted in \cref{fig:setting}.
The moving window allows to reduce the total simulation domain to just a region around the solidification front.
Typical simulations in representative volume elements may require between $10\,000$ to $20\,000$ compute cores for multiple hours \cite{Hoetzer2015}.
A typical phase-field simulation of the directional solidification of the ternary eutectic system \systemAlAgCu{} is shown in \cref{fig:pf-sim}.
The $12\,000\times12\,000\times65\,142$ voxel cell domain is calculated on the SuperMUC system with  $19\,200$ cores within approximately $19$ hours.

Based on this highly parallel and optimized solver, the eutectic solidification of idealized systems and real ternary alloys like  \systemAlAgCu{} were investigated in large scale domains and the experimentally assumed growth of spirals could be demonstrated \cite{hoetzer15c,Hoetzer2015}.

\section{Benchmarks}\label{sec:benchmarks}

In this section, we present benchmarks to illustrate the performance of our checkpointing implementation. We discuss different aspects of the checkpointing scheme, explain and present the respective performance results.

\begin{table*}
\centering
\footnotesize
\renewcommand{\arraystretch}{1.2}
\begin{tabularx}{\textwidth}{ X | X | X | X | X }
& \textit{LSS cluster} \newline (LSS at FAU Erlangen) & \textit{Emmy cluster} \newline (RRZE at FAU Erlangen) & \textit{SuperMUC \newline Phase I, Thin Nodes} \newline (Leibniz Supercomputing \newline Centre Garching) & \textit{JUQUEEN} \newline (J\"ulich Supercomputing \newline Centre (JSC))\\
\hline\hline
processor type (cores) & Xeon E7-4830 (8) & Xeon 2660v2 (10) & Xeon E5-2680 (8) & IBM PowerPC A2 (16) \\
base frequency &  2.13 GHz & 2.20 GHz & 2.70 GHz & 1.60 GHz \\
compute nodes & 8 & 560 & 18 (islands) $\times$ 512 (nodes) & 28\,672 \\
processors per node & 4 & 2 & 2 & 1 \\
memory per node & 256 GB & 64 GB & 32 GB & 16 GB \\
interconnect technology & InfiniBand QDR & InfiniBand QDR & InfiniBand FDR10 & 5D Torus - 40 GBps \\
MPI & OpenMPI 1.7 \newline (ULFM Fork) & Intel MPI 5.1 & IBM MPI 1.4 & MPICH 2 \\
\hline
\end{tabularx}
\vspace{6pt}
\caption{Test systems that are used for the benchmarks. The LSS cluster is mainly used to test the runtime fault tolerance of the implementation by installing ULFM. On the SuperMUC system, the intra-island topology is based on a non-blocking tree while the inter-island topology is only based on a 4:1 pruned tree. Therefore the connection quality is expected to decrease when moving from one to multiple islands.}
\label{tab:systemspecs}
\end{table*}

\subsection{The test cases}

To benchmark the checkpointing scheme of \cref{sec:implementation}, we simulate the directional solidification of a ternary eutectic system using the implementation presented in \cite{Bauer2015} of the phase-field model  introduced in \cref{sec:phasefield}. 
For the simulation parameters, the values of \cite{hoetzer15c} to study spiral growth are used.
To resolve spiral growth, large domain sizes and millions of time steps are required, resulting in massively parallel and long-running simulations. Since we expect such large-scale simulations to be subject to hardware failures in the future, the scenario is a relevant example to demonstrate resilience techniques.
However, as noted in \cref{sec:recoverystrategies}, the checkpointing algorithm is not bound to a specific algorithm and can therefore also be applied to other applications using different data structures. In this sense, the block data items that are exchanged among the processes are black-boxes to the implementation. They solely need to implement respective serialization and deserialization routines. 

\hl{The moving window technique} \cite{Vondrous2014} depicted in \cref{fig:setting} is realized by adding information about the current absolute coordinates of the domain to the block data. By simply adding this data to the checkpoint, there is no additional work involved but implementing the serialization and deserialization routines for the cell coordinates.

The domain is set to a fixed size in the $z$-direction while different sizes are employed in the $x$- and $y$-directions. This is achieved either by varying numbers of cells per block or by varying the number of blocks in the respective directions.
The model requires $12$ floating point values per cell.

The tests are performed on the clusters listed in \cref{tab:systemspecs}.

\subsection{System Size and Scaling}\label{sec:scaling}

For the proposed checkpointing scheme the amount of data to be exchanged between processes during a checkpoint depends on the redundancy and not on the overall number of employed processes. If we choose the number of copies that are backed up to be constant, the strategy scales, i.e. the duration of the checkpointing step stays constant and independent of the number of processes.

To demonstrate the scalability of the checkpointing scheme, we set up a weak scaling experiment \cite{padua2011encyclopedia} by fixing the number of blocks per process while increasing the size of the domain and therefore the number of processes. We perform the weak scaling with a different number of cells per block in multiple benchmarks to see how our implementation scales depending on the amount of transferred data. The domain size is kept constant in the $z$-direction. As the number of blocks is not always an integer multiple of the number of processes, we also plot the average number of blocks per process if it is not constant in all runs.

\begin{figure}[t]
\centering
\includegraphics[width=1.0\columnwidth]{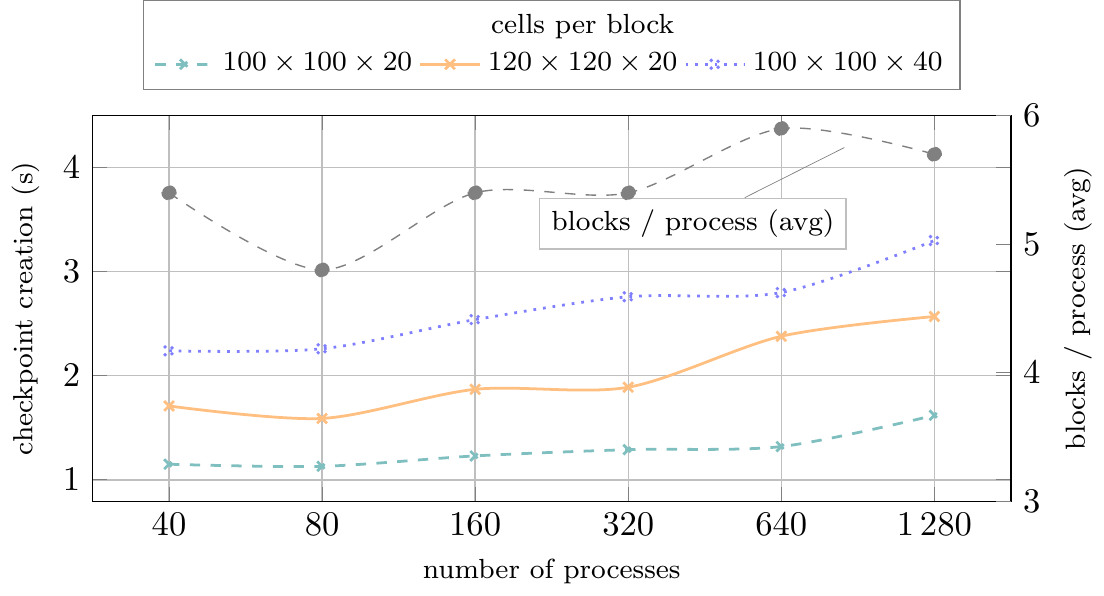}
\caption{Weak scaling of the checkpointing duration on the Emmy cluster with different block sizes. The average number of blocks per process varied slightly for different numbers of processes.}
\label{fig:weakscalingemmy}
\end{figure}

We also perform the experiment on different architectures. The results from the Emmy cluster, where we use up to $64$ nodes with up to $1\,280$ \mpi{} processes in total can be found in \cref{fig:weakscalingemmy}. 

\tikzset{
	every pin/.style={draw=gray!30, fill=white,rectangle,rounded corners=3pt,font=\tiny}
}

\begin{figure}[h]
\centering
\includegraphics[width=1.0\columnwidth]{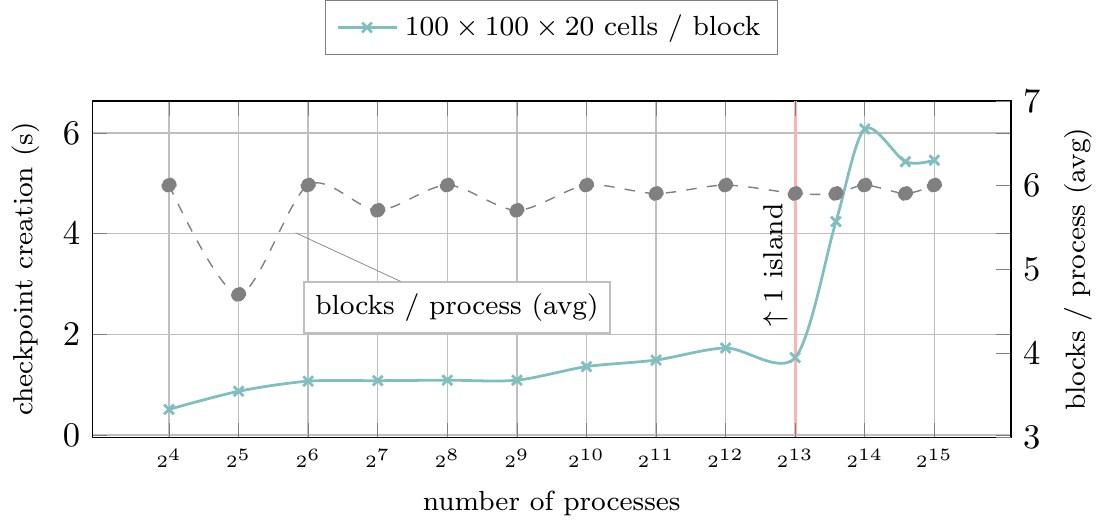}
\caption{Weak scaling of the checkpointing duration for a block size of $100 \times 100 \times 20$ on the SuperMUC system. The average number of blocks per process varied slightly for different numbers of processes.}
\label{fig:weakscalingsupermuc}
\end{figure}

On the SuperMUC system, we perform the benchmark on up to $2\,048$ nodes, with $32\,768$ MPI processes for the largest scenario. The results can be found in \cref{fig:weakscalingsupermuc}.

\begin{figure}[h]
\centering
\includegraphics[width=1.0\columnwidth]{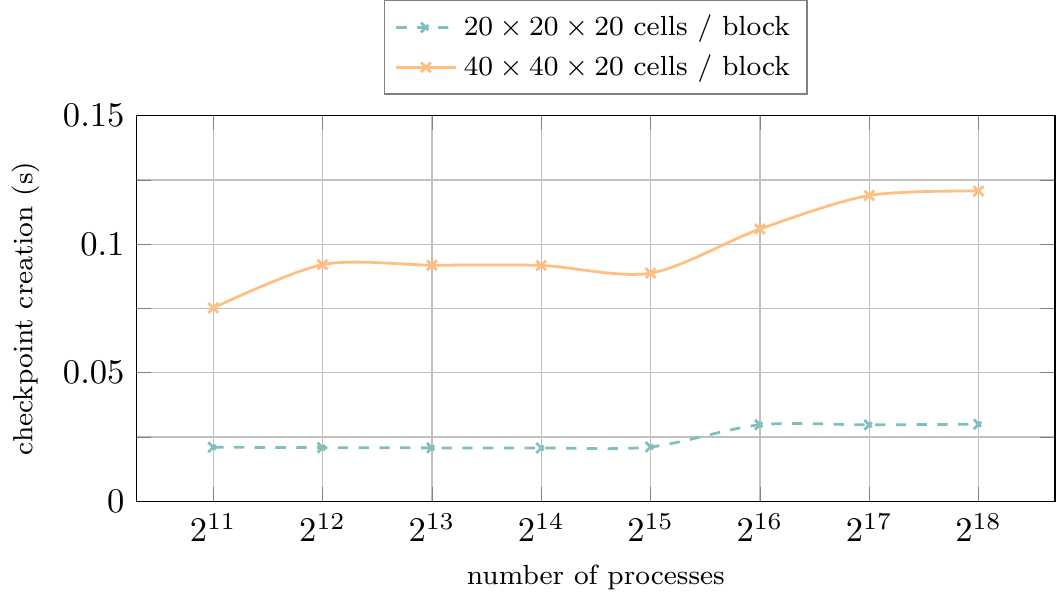}
\caption{\hl{Weak scaling of the checkpointing duration for different block sizes on the JUQUEEN system. The number of blocks per process is fixed to $1$.}}
\label{fig:weakscalingjuqueen}
\end{figure}

\hl{In a third run, we test our implementation on JUQUEEN} \cite{Juqueen} \hl{with up to $16\,384$ nodes and $262\,144$ MPI processes, see} \cref{fig:weakscalingjuqueen}.

\hl{All three} benchmarks show, that the algorithm scales almost perfectly, i.e. independently of the number of employed processes. The time increases only slightly with the increasing number of processes. In the SuperMUC scaling we can see a sudden increase in the checkpointing duration. This is expected since the simulation spans to multiple islands from that point on. As in our setup the pair-wise checkpointing strategy locates the backups of a process pair on two different islands, the slower data exchange increases the duration of the checkpoint creation. This can be avoided by altering the distribution strategy or by pinning the \mpi{} ranks in a proper order to force that no backups are sent to a different island. On the other hand, having the backup pairs on different islands guards against failure of an entire island. With increasing number of islands (two islands for $2^{14}$ processes, four islands for $2^{15}$ processes) the checkpointing duration stays stable. 

\hl{In the scalings on Emmy and JUQUEEN, we additionally compare the checkpointing duration among simulations with different numbers of cells per block. Comparing for example the block sizes $100 \times 100 \times 20$ and $100 \times 100 \times 40$ in} \cref{fig:weakscalingemmy} \hl{(the latter containing twice as many cells and therefore twice as much data) we see that the checkpointing duration almost exactly doubles and therefore seems to depend directly proportional on the amount of data that has to be transferred.}


\subsection{Overhead}

Although, in theory, the checkpointing duration scales perfectly this does not necessarily mean its time consumption is negligible. Recall the approximation to the theoretical optimal checkpointing frequency $f_\text{\textsc{opt}}$ from \eqref{eq:optimalcheckpointingfrequency}.

In fact, the overall time that should be spent on checkpointing according to the approximated optimal checkpointing frequency depends indirectly on the number of processes in the system as this number is related to the MTBF, see \eqref{eq:mtbfparallel}.

When the MTBF $\mu$ decreases, the optimal checkpointing frequency $f_\text{\textsc{opt}}$ will increase. Thus, although the checkpointing duration itself stays constant, the frequency and therefore the overhead will grow with larger systems if the optimal checkpointing frequency is employed.

Not only the frequency, but also the size of the data influences the checkpointing duration. This means, larger per-process workloads are also expected to increase the time spent checkpointing. However, an increased absolute checkpointing duration also increases the optimal checkpointing interval.

\begin{figure}[h]
\centering
\includegraphics[width=1.0\columnwidth]{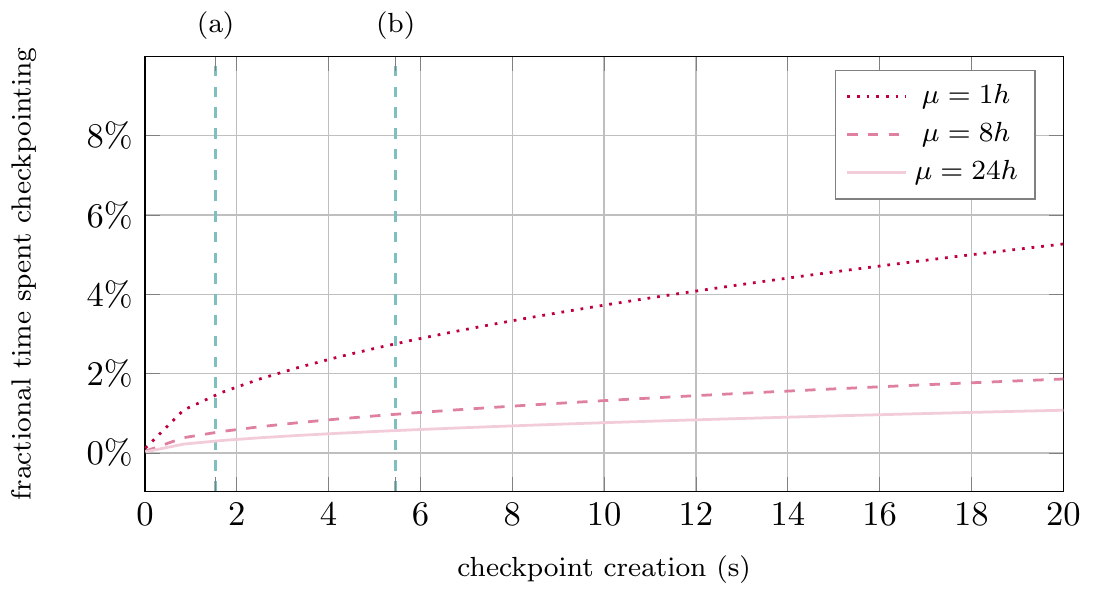}
\caption{Graph of the overhead due to checkpointing using an approximation to the optimal checkpointing frequency. The overhead function depends on the MTBF of the system $\mu$ and the checkpointing duration $C$. (a) and (b) mark the checkpointing time for the SuperMUC scenario of \cref{fig:weakscalingsupermuc} for $2^{13}$ and $2^{15}$ processes respectively.}
\label{fig:wastegraph}
\end{figure}

To get an estimate for the overhead that is introduced when the checkpointing frequency is set to the theoretical optimal value in \eqref{eq:optimalcheckpointingfrequency}, we use the measurements from the SuperMUC system. We then compute the overhead by dividing the respective checkpointing duration $C$ by the approximation to the optimal checkpointing interval $T_\text{\textsc{opt}} = 1 / f_\text{\textsc{opt}}$:
\begin{equation}
\text{\textsc{Overhead}} = \frac{C}{T_\text{\textsc{opt}}} = \frac{C}{\sqrt{2 \mu C}} \text{.}
\end{equation}
In \cref{fig:wastegraph} we plot the function for different hypothetical MTBFs $\mu$ and using measured values for the checkpointing duration $C$.

With the approximate optimal checkpointing frequency, the checkpointing overhead lies below $3 \%$ even if the MTBF is only one hour.

\subsection{Recovery}

Since the block data that needs to be restored during the recovery already resides on the respective processes, the recovery does not involve any inter-process communication (as illustrated in \cref{fig:snapshotstrategy}). The time is only spent on deserialization of the data. As the data resides in main memory, we do not expect the time spent during the recovery process to have a large impact on the runtime. \hlgreen{Still, a load balancing step is in general necessary to maintain the performance of the application after a failure.}

To measure the time it takes to recover the block data, we simulate a failure by erasing the block data from all processes and forcing them to restore the block data of the partner process from the last checkpoint. This way, each process is then responsible for the block data of another process. No process is actually killed during this procedure. 

\begin{figure}[h]
\centering
\includegraphics[width=1.0\columnwidth]{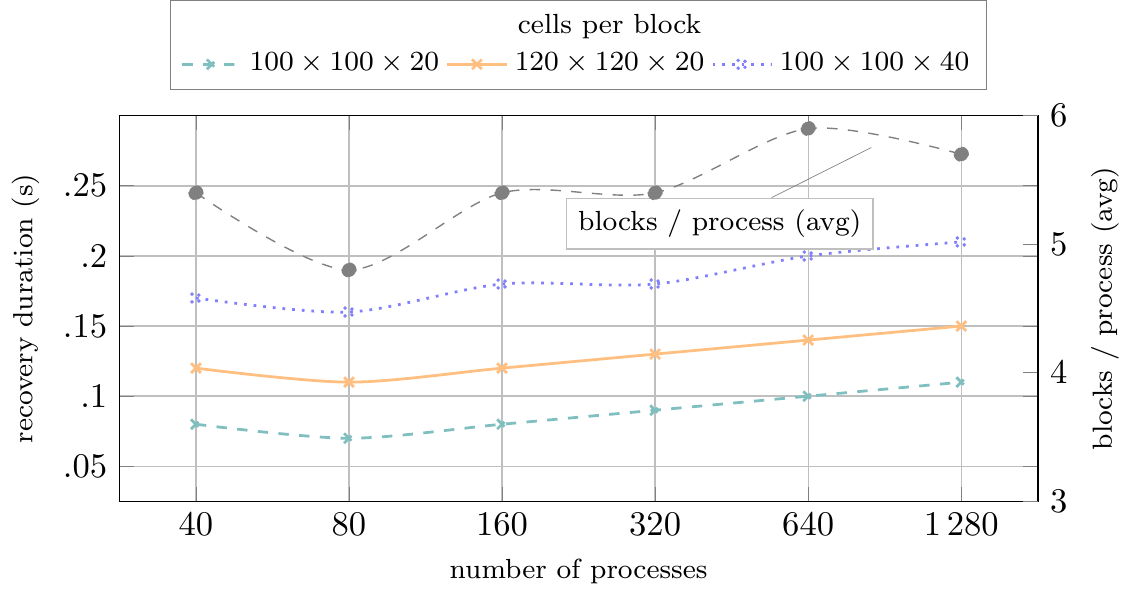}
\caption{Weak scaling of the recovery duration for different block sizes on the Emmy cluster. The average number of blocks per process slightly varied for different numbers of processes due to the test setup.}
\label{fig:recoveryemmy}
\end{figure}

The benchmark is performed on the Emmy cluster and the results shown in \cref{fig:recoveryemmy}.

The recovery scales almost perfectly since there is no communication involved and it only takes some milliseconds to restore the data. \hl{Also, analogously to the observation in} \cref{sec:scaling} \hl{we observe the time to recover the snapshot to be directly dependent on the amount of data that was backed up. Compared to the runtime of a typical simulation, the time that is spent on recovery itself - excluding additional load balancing - can be neglected.}

\hl{Nevertheless, the load imbalance that is introduced by the fault must be tackled in a second step. A systematic study of the load balancing mechanisms in \walberla{} can be found in} \cite{Schornbaum2017}.

\subsection{Fault Tolerance}

To test the resilience of the application, we install the OpenMPI ULFM implementation on both, desktop machines and the LSS cluster at the Chair for System Simulation in Erlangen and simulate process faults via kill signals from within the application (using the C library function \lstinline{raise()}) and from the outside (using the \lstinline{kill} command).

\begin{figure*}[h]
\centering
\begin{subfigure}{0.3\textwidth}
\includegraphics[width=0.8\textwidth]{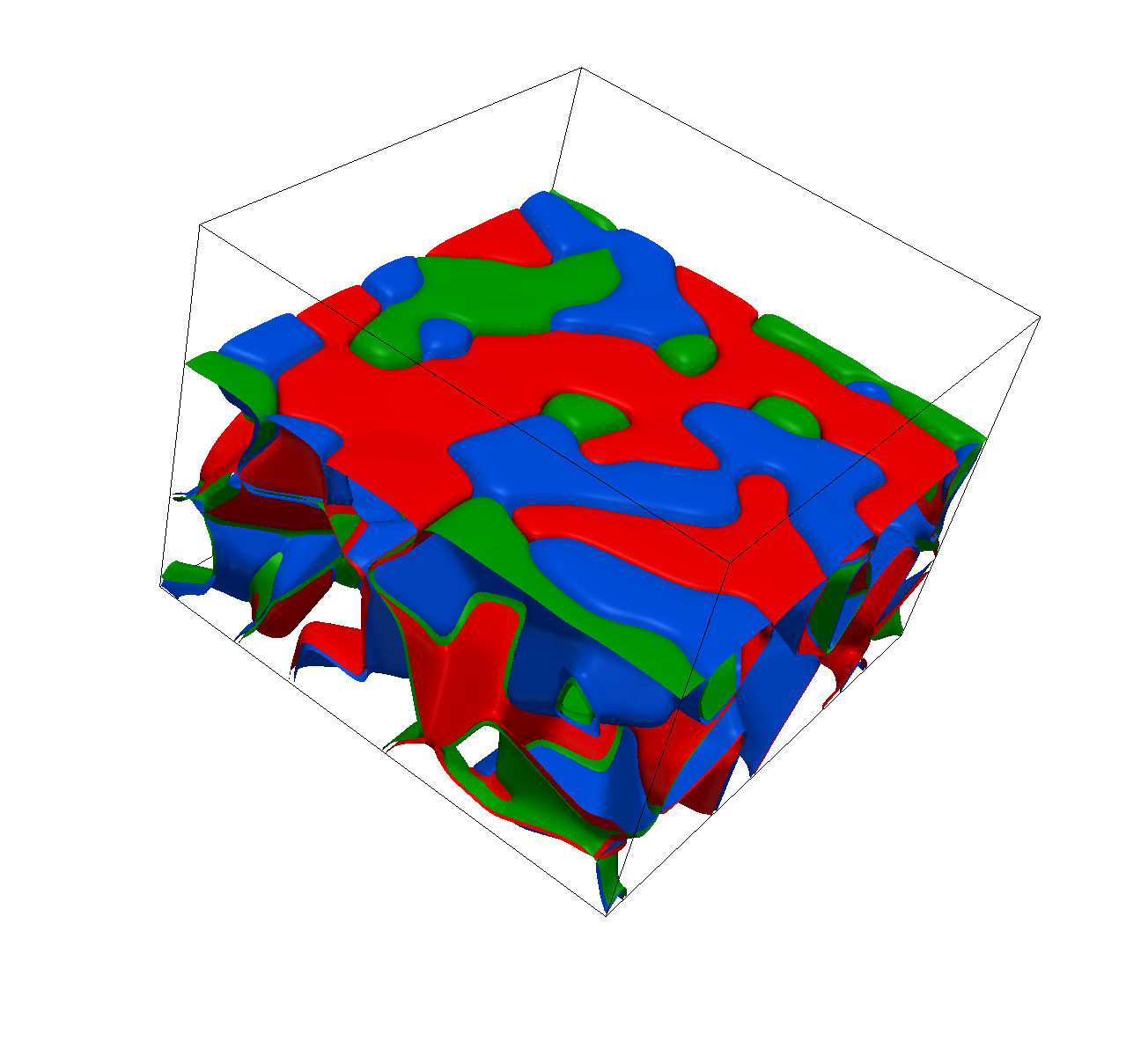}
\caption{time step $2\,500$}
\end{subfigure}
\begin{subfigure}{0.3\textwidth}
\includegraphics[width=0.8\textwidth]{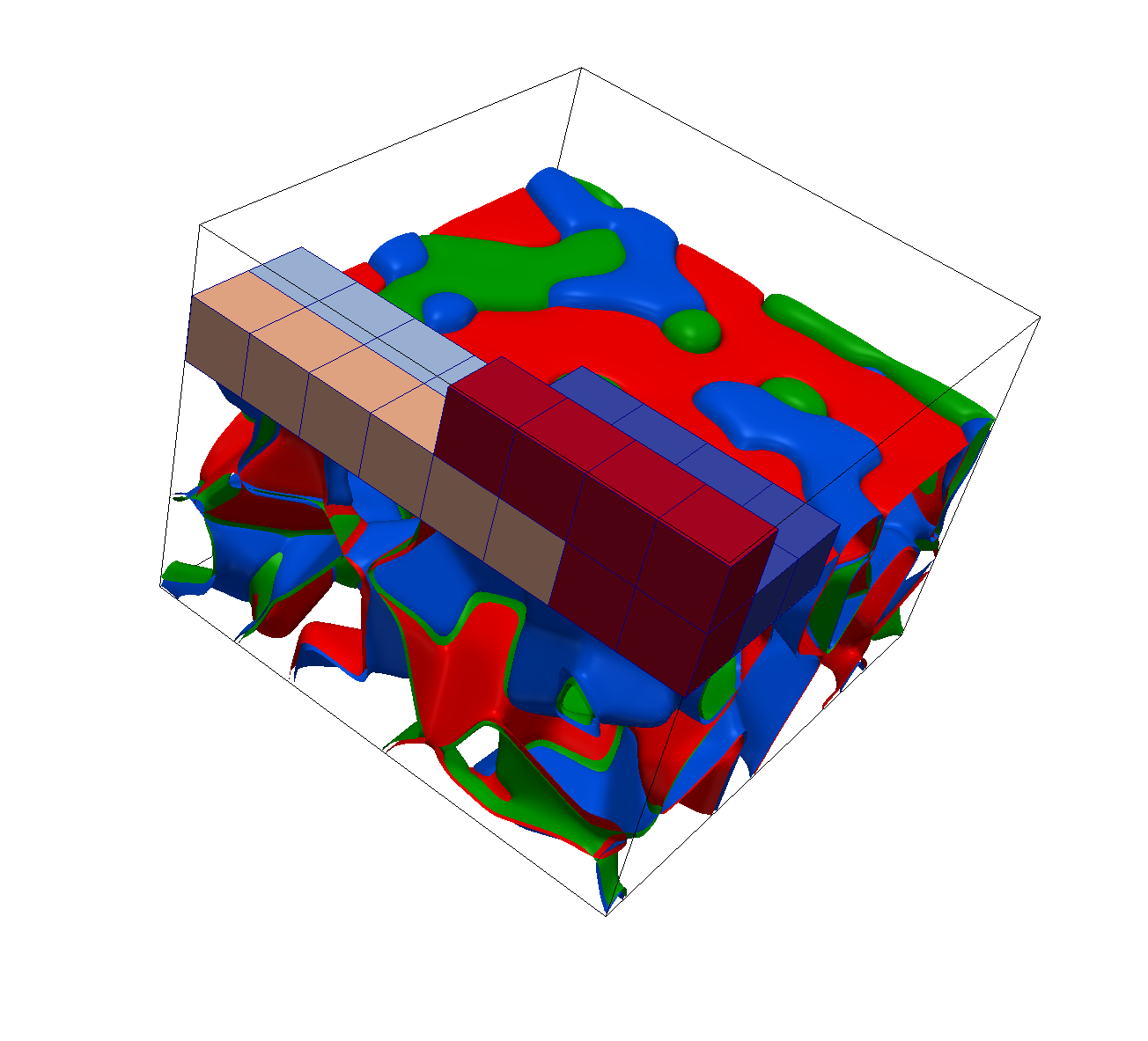}
\caption{time step $5\,600$}
\label{fig:phasefieldfailure}
\end{subfigure}
\begin{subfigure}{0.3\textwidth}
\includegraphics[width=0.8\textwidth]{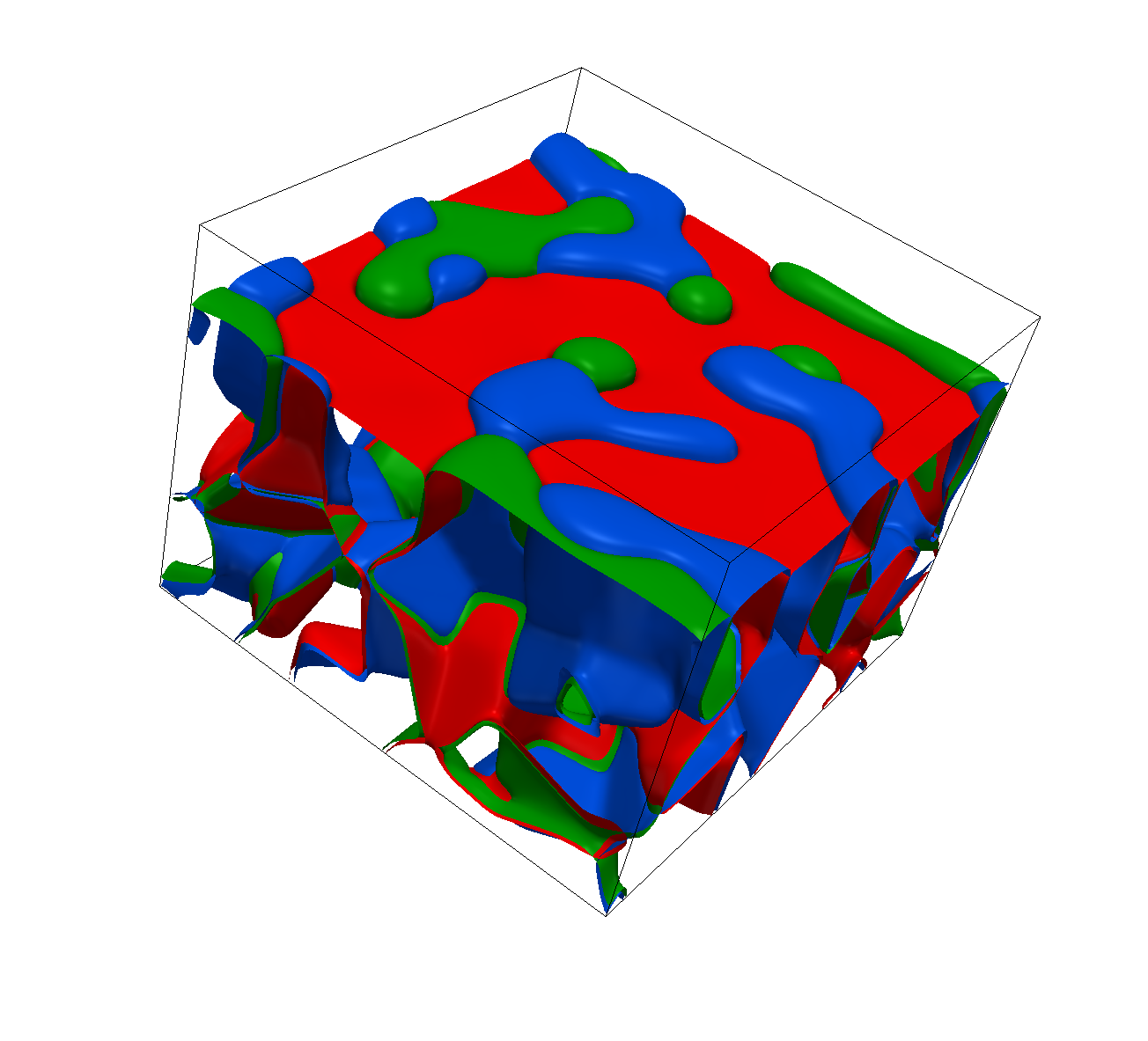}
\caption{time step $9\,950$}
\end{subfigure}
\caption{Phase-field simulation of the direction solidification of the ternary eutectic system \systemAlAgCu{} in a 160 $\times$ 160 $\times$ 120 voxel cell domain, run with 64 \mpi{} processes on two nodes of the LSS cluster. During the simulation we send kill signals to four \mpi{} processes. The affected blocks are highlighted in \cref{fig:phasefieldfailure}. Blocks that reside on the same process are equally colored. After the processes fault, the simulation restores the last checkpoint during runtime and then proceeds.}
\label{fig:faulttolerancephasefield}
\end{figure*}

\Cref{fig:faulttolerancephasefield} illustrates a test on the LSS cluster where we simulate the directional solidification using the ULFM \mpi{} implementation. In the depicted scenario, each process carries six blocks with $20^{3}$ cells per block. During the simulation we send kill signals to four \mpi{} processes using the \lstinline{kill} command. The application then recovers from the process faults during runtime and successfully restores the last snapshot of the simulation data using the pair-wise checkpointing scheme as described in \cref{sec:implementation}. This way the simulation is continued without halting and restarting the application.

It should be noted, that the OpenMPI fork that implements ULFM is still experimental \cite{ULFM2013,Bland2015}. Still, the fork serves as a tool to provide a proof of concept implementation - regarding the MPI library itself as well as applications using it.

\subsection{Flexibility of the concept}

\begin{figure*}[h]
\centering
\begin{subfigure}{\textwidth}
\includegraphics[width=1.0\columnwidth]{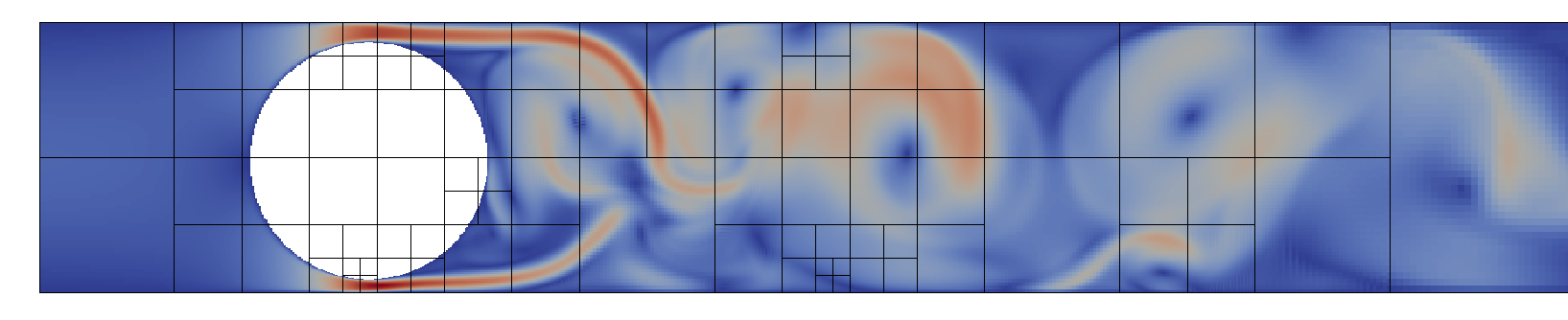}
\caption{\hl{velocity magnitude (after $40\,000$ iterations)}}
\label{fig:lbmvelocity}
\end{subfigure}

\begin{subfigure}{\textwidth}
\includegraphics[width=1.0\columnwidth]{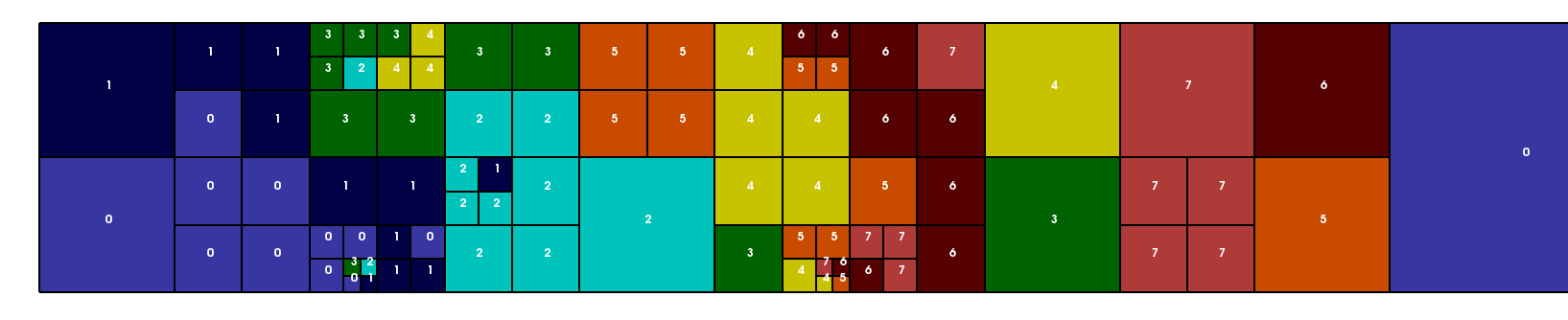}
\caption{\hl{domain partitioning of healthy run (after $40\,000$ iterations) - process ranks indicated via color / number}}
\label{fig:lbmpartitioninghealthy}
\end{subfigure}

\begin{subfigure}{\textwidth}
\includegraphics[width=1.0\columnwidth]{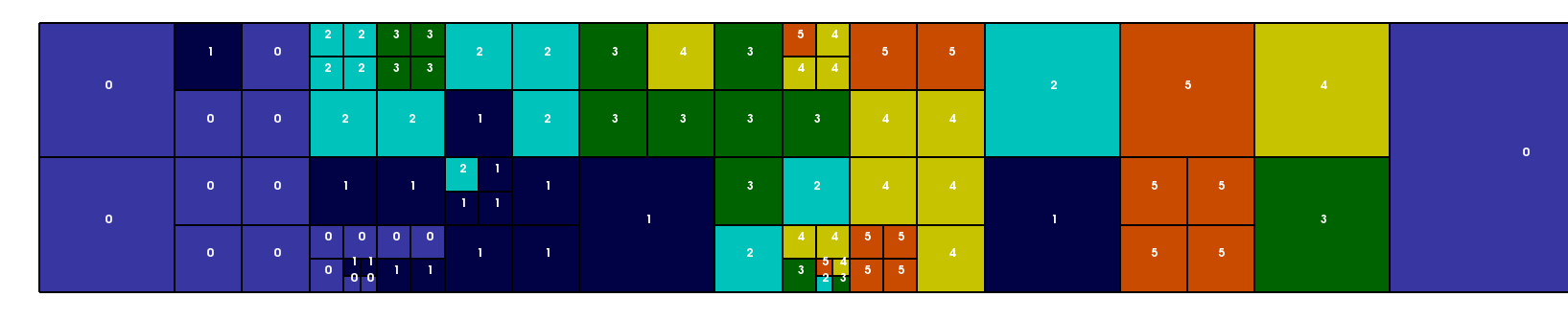}
\caption{\hl{domain partitioning of faulty run (after $40\,000$ iterations) - process ranks indicated via color / number}}
\label{fig:lbmpartitioningfaulty}
\end{subfigure}
\caption{\hl{Two-dimensional Lattice Boltzmann simulation of laminar flow around a cylinder run with the \walberla{} framework. In }\cref{fig:lbmvelocity}\hl{ the velocity magnitude of the fluid is depicted together with an adaptively refined block structure. Each block contains $40 \times 40$ cells. }\Cref{fig:lbmpartitioninghealthy}\hl{ shows the assignment of the blocks to the eight employed processes. In a second run, two processes (rank 6 and 7, colored in red and darker red) are killed during the simulation (after roughly $34\,000$ iterations). The resulting load distribution after the recovery and load balancing among the remaining processes is visualized in }\cref{fig:lbmpartitioningfaulty}.}
\label{fig:faulttolerancelbm}
\end{figure*}

\hl{To demonstrate the flexibility of our implementation, we also apply the presented checkpoint-rollback technique to a lattice Boltzmann simulation} \cite{Godenschwager2013}. \Cref{fig:faulttolerancelbm}\hl{ shows a LBM simulation of laminar fluid flow around a cylinder in two dimensions using adaptive mesh refinement on eight processes} \cite{Schornbaum2015,Schornbaum2017}\hl{. Starting with a regular grid, the block structure is successively refined to obtain better resolutions in certain subdomains. Each block contains a constant number of cells ($40 \times 40$ in this scenario). Below the velocity magnitude and the overlayed domain partitioning, in} \cref{fig:lbmpartitioninghealthy}\hl{ we can see the domain partitioning together with the block distribution to the eight employed processes. In a second run, two processes (rank 6 and 7) are killed during the simulation (after roughly $34\,000$ iterations). The respective block distribution is shown in} \cref{fig:lbmpartitioningfaulty}. \hl{Right after the fault, the load is re-balanced among the surviving processes.}

\section{Conclusion}

In this work, we demonstrate how a simulation framework can be augmented by resilience features while maintaining performance and scalability. We propose a distributed and diskless checkpointing algorithm that can be combined with the runtime recovery mechanisms of the ULFM MPI extension to provide a fast and scalable method to implement resilience in modern applications. Using a double-buffer model, we make the checkpointing procedure itself resilient to process failures.

Applying the approach to the simulation of phase-field problems in natural science, we show that it only introduces minor runtime overhead and scales almost perfectly. \hl{In the benchmark scenario with the largest domain consisting of about $40$ billion cells and more than ten times as many floating point values, the checkpoint creation takes less than $7$ seconds using $2^{15}$ MPI processes. We also show that our implementation scales up to $2^{18}$ MPI processes.} Since the recovery phase does not require inter-process communication, it scales and takes less than a second in all performed benchmarks. Using an approximation to the optimal checkpointing frequency, we show that the checkpoint creation only creates an overhead of less than $4 \%$ of the application's runtime in a theoretical system with a MTBF of one hour.

In future work, we will employ the proposed mechanisms to problems with different data structures. Further, we plan to improve the strategy by employing data compression or more sophisticated distribution algorithms.

\section{Acknowledgements}
The authors gratefully acknowledge funding by the joint BMBF project \skampy. The authors gratefully acknowledge the Gauss Centre for Supercomputing e.V. (\url{www.gauss-centre.eu}) for funding this project by providing computing time on the GCS Supercomputer JUQUEEN at Jülich Supercomputing Centre (JSC) and SuperMUC at Leibniz Supercomputing Centre (\url{www.lrz.de}).

\bibliographystyle{apalike}
\bibliography{bibliography}

\end{document}